\documentclass[acmlarge]{acmart}
\makeatletter
\newcommand{\confshort}{\acmConference@shortname}
\newcommand{\conffull}{\acmConference@name}
\newcommand{\confdate}{\acmConference@date}
\newcommand{\confloc}{\acmConference@venue}
\AtBeginDocument{
  \fancypagestyle{firstpagestyle}{
    \fancyhead{}%
    \fancyfoot[C]{}%
  }
  \fancyhf{}
  \fancyhead[LO]{\@headfootfont\shorttitle}%
  \fancyhead[RE]{\@headfootfont\@shortauthors}%
  \fancyhead[LE]{\@headfootfont\footnotesize \confshort, \confdate, \confloc}%
  \fancyhead[RO]{\@headfootfont\footnotesize \confshort, \confdate, \confloc}%
  \fancyfoot[C]{}%
}
\makeatother
\acmBooktitle{\conffull\@ (\confshort), \confdate, \confloc}

\AtBeginDocument{%
  }

\usepackage[inkscape=false]{svg}
\usepackage{multirow}
\usepackage{subcaption}
\usepackage[percent]{overpic}
\usepackage{longtable}
\usepackage{tcolorbox}

\copyrightyear{2026}
\acmYear{2026}
\setcopyright{cc}
\setcctype{by-nc-nd}
\acmConference[FAccT '26]{The 2026 ACM Conference on Fairness, Accountability, and Transparency}{June 25--28, 2026}{Montreal, QC, Canada}
\acmBooktitle{The 2026 ACM Conference on Fairness, Accountability, and Transparency (FAccT '26), June 25--28, 2026, Montreal, QC, Canada}
\acmDOI{10.1145/3805689.3812381}
\acmISBN{979-8-4007-2596-8/2026/06}

\begin{document}

\title[Resume Screening, Fast and Slow]{Resume Screening, Fast and Slow: (Biased) AI Recommendations' Influence on Human Decision Making}

\author{Kyra Wilson}
\email{kywi@uw.edu}
\orcid{0009-0001-3955-3409}
\affiliation{%
  \institution{University of Washington}
  \city{Seattle}
  \country{USA}
}

\author{Mattea Sim}
\affiliation{%
  \institution{Georgetown University}
  \city{Washington, D.C.}
  \country{USA}}

\author{Anna-Maria Gueorguieva}
\affiliation{%
  \institution{University of Washington}
  \city{Seattle}
  \country{USA}}

\author{Soham Chatterjee}
\affiliation{%
  \institution{International Institute of Information Technology}
  \city{Hyderabad}
  \country{India}}

\author{Aylin Caliskan}
\affiliation{%
  \institution{University of Washington}
  \city{Seattle}
  \country{USA}}

\renewcommand{\shortauthors}{Wilson et al.}

\begin{abstract}
    
    AI is increasingly being used collaboratively with people to make decisions in high-stakes domains, but this new paradigm is still not well-understood in many respects---particularly regarding how AI that replicates human social biases influences people's decision making processes and how that can influence outcomes. In this study, we analyzed the time people spend viewing candidate resumes from an experiment investigating biased AI resume screening to evaluate decision-making fairness and cognitive processes underlying human-AI collaboration. We found that spending more time viewing resumes corresponds to candidates' selection chance increasing by 3-4\% if they are not recommended, and people may spend up to 55.6\% longer viewing resumes when no AI recommendations are given. Furthermore, people who completed an implicit association test (IAT) before resume screening were significantly more likely to evaluate candidates of different races for the same amount of time, and their IAT scores were also predictive of the time spent in human-AI collaboration. These results demonstrate how people's decision-making processes can be insufficient for overseeing AI in high-stakes domains.

\end{abstract}



\maketitle

\section{Introduction}

As hiring decisions are increasingly made collaboratively between humans and AI systems using “human-in-the-loop” AI teaming (AI-HITL) setups \citep{resumebuilder}, there is growing interest in the societal benefits and/or harms that could arise. For example, AI recommendations have been used to reduce recruiting time by 75\% for one company, leading to a savings of over one million dollars \citep{hirevue}. Alternatively, these systems can also exhibit preferences for candidates based on their social identities rather than qualifications (\textit{bias}) \citep{wilson2024gender} as happened in 2018 when Amazon developed a system which discriminated against women \citep{dastin2018}. One further risk is that humans may also be ineffective at detecting and counteracting these biases prior to making decisions \citep{wilson2025no}, leading to outcomes which are undesirable, or in some cases, illegal. Therefore, using AI-HITL systems which can improve the efficiency and quality of decision making while also not reducing people's ability to oversee and correct critical failures is essential to ensuring that these models do not cause serious harm in the world. Achieving this will require better understanding of people's interactions with biased AI (beyond the outcomes) to inform future development and oversight strategies that encourage deep, analytical thinking which is able to overcome AI biases or errors \citep{smith2000dual, payne2005conceptualizing}.

In this study, we investigate under what conditions people are able to engage in more analytical processing in the presence of racially biased AI recommendations for resume screening. While related work has investigated collaboration between people and AI for hiring \citep{wilson2025no}, we are the first (to our knowledge at the time of writing) to specifically focus on how \textit{biased} AI impacts interactions in this domain at the level of human decision-making processes rather than human-AI decision-making outcomes in isolation. This enables us to connect (unfair) outcomes to the (potentially flawed and heuristic) processes that led to them and understand how to increase decision-making efficiency while maintaining quality of outcomes. To do this, we analyze the data collected by \citet{wilson2025no} to measure the effects of racially biased AI recommendations on people's selection rates of racially diverse candidate resumes. This large-scale human subjects experiment contained responses from over 500 participants in 1,500 AI-HITL resume-screening scenarios comparing Asian, Black, Hispanic, and white candidates for 16 high and low status occupations. They investigated five different conditions of AI bias that varied in the candidates preferred and the magnitude of preference, as well as conditions without AI recommendations. They found that in this final condition, people selected candidates of different races at approximately equal rates, and when AI recommendations were racially biased, people's final candidate selections mirrored those biases. People were only marginally successful at mitigating AI bias when it was extremely severe (e.g., candidates recommended entirely based on race), which left open questions as to what made people's oversight of AI decision making more effective in some circumstances than others.

To address this, we analyzed people's decision-making process when interacting with various kinds of biased AI recommendations in the resume-screening task conducted by \citet{wilson2025no}. While there are numerous facets of decision making \citep{phillips2016thinking}, we specifically focused on how long people spent viewing resumes based on factors like races of the presented candidates, biases of AI recommendations, and higher-level system or individual characteristics. There are a number of reasons resume viewing duration is worth further study: first, it can be considered a proxy to quality of resume screening \citep{backstrom2017increasing}. If people spend different amounts of time viewing candidates of different races in certain conditions (i.e., candidates do not receive the same quality of evaluation), that can be one indication of procedural unfairness (inequality in procedures leading to a decision) in the resume-screening process leading to distributive unfairness in outcomes (equal racial representation among the final selected candidates) \citep{rauh2024gaps, barocas-hardt-narayanan}. Second, time can be informative of underlying cognitive processes which rely more on heuristic or associative reasoning rather than deep, analytical reasoning \citep{phillips2016thinking}. While many studies of decision making in hiring emphasize the role of quick, intuitive thinking \citep{miles2014recruitment, simon1987making}, deeper reasoning is also needed to counteract stereotypes and bias \citep{smith2000dual, payne2005conceptualizing}. Finally, as increased efficiency is one major driver of AI adoption in high-stakes domains which impact people's economic security and well-being \citep{hirevue}, analyzing people's time spent in these tasks enables a clearer evaluation of efficiency vs. quality tradeoffs.

This analysis\footnote{Data and scripts for reproduction are available at \url{https://github.com/kyrawilson/Resume-Screening-Fast-and-Slow}.} to extend the results of \citet{wilson2025no} is summarized by the following research questions and contributions:

\begin{itemize}
    \item RQ1: How does the interaction of AI recommendations and the time people spend evaluating resumes influence candidates' likelihood of selection?  $\rightarrow$ \textbf{People spend 27 seconds on average evaluating each resume; negative recommendations were associated with candidates' selection chance increasing 3-4\% for every additional 30 seconds spent viewing their resume. For candidates who were recommended, increased viewing duration was associated with a smaller, but still significant, decreased chance of selection.} 
    \item RQ2: Do (racially biased) AI recommendations affect the amount of time that people spend viewing and evaluating white vs. non-white candidates? $\rightarrow$ \textbf{Not having AI recommendations is associated with 55.6\% longer durations viewing resumes, regardless of candidates' race; otherwise, significant differences emerged when AI recommendations were extremely severe and incongruent with common stereotypes and participants were evaluating high status jobs. In this case, participants spent 20.1\% less time evaluating non-white candidates than white candidates.} 
    \item RQ3: Does raising people's awareness of race-status associations before they interact with biased AI recommendations affect amount of time that people spend viewing and evaluating white vs. non-white candidates? $\rightarrow$ \textbf{The time people spent evaluating Black vs. white candidates for high status occupations was not significantly different if they took an implicit association test (IAT) first, indicating a change in decision-making processes.} 
    \item RQ4: Do people's implicit beliefs about race and status affect the amount of time they spend viewing and evaluating white vs. non-white candidates when interacting with biased AI recommendations? $\rightarrow$ \textbf{Race-status associations did not have a uniform effect across racial groups, but stronger stereotypical race-status associations consistently increased the time participants spent reading Black and white candidates' resumes regardless of the bias of AI recommendations.} 
\end{itemize}

\section{Related Work}

AI systems are being used as decision-making support tools in a wide variety of domains, and there has been a significant amount of research into how AI usage impacts people's decision making and its outcomes. For example, some research investigates AI-HITL through the lens of productivity, showing that even though users perceive AI as improving productivity, they actually spend more time on tasks when using AI \citep{qian2024take}. Other work has emphasized the importance of developing AI systems which complement the existing knowledge of human decision makers to maximize overall system performance \citep{hemmer2025complementarity, reverberi2022experimental}. Among this body of research, key results suggest that individual factors, such as people's motivation to engage in effortful cognitive activities \citep{buccinca2021trust, eisbach2023optimizing}, perceptions of the utility of AI recommendations \citep{lai2019human}, or decision-making styles \citep{mei2025passing} moderate the success rates of AI-HITL collaboration.

One limitation of these studies, however, is that they have not investigated how people's individual biases moderate their interactions with biased AI models. This is especially important given that AI tools used for high-stakes tasks such as hiring can be biased against groups of people based on race, gender, or disability \citep{glazko2024identifying, wilson2024gender}. The studies which have investigated interactions with biased AI in the hiring domain shown that people are often not aware of model biases \citep{kuhl2025bias, rosenthal2024michael} and follow biased recommendations even among equally qualified candidates \citep{wilson2025no}. However, these studies have been limited to analyzing the outcomes of hiring decisions but not the decision-making processes leading to them, which can also be informative for system design and harm mitigation.

While there are numerous psychological theories of decision making, many of them can be summarized as a dual-process model \citep{oxford_dual, kahneman2013perspective, Stanovich_West_2000}. In its simplest form, this model posits that System 1 is responsible for making decisions which are fast; based on heuristics, associations, or biases; and not guaranteed to be correct. System 2, on the other hand, is slower, requires more effortful analytical and rational processing, and is more likely to reach the correct conclusion. Many socio-cognitive processes including stereotyping \citep{devine1989stereotypes}, persuasion \citep{petty1986elaboration}, and moral judgment \citep{haidt2001emotional} have been explained using a dual-process model. Furthermore, a number of studies have also found that the effectiveness of AI-assisted decision making can be influenced by the activation of System 1 or 2 thinking and find that over-reliance is more likely when the latter is not activated \citep{buccinca2021trust, chen2023understanding, rastogi2022deciding}.

People's own beliefs and biases may also influence their decision making, whether they be conscious or unconscious \citep{ajzen2015explicit}. The latter are typically measured using IATs, first proposed by \citet{greenwald1998measuring} as a way to measure associations via differences in reaction times when sorting words or pictures representing two concepts of interest (e.g., gender/occupation, race/status). Most studies predicting discriminatory decision-making behavior with IATs use tests associating social categories and valence; however, associating categories with beliefs may be better at predicting behavior \citep{rudman2007discrimination, montgomery2024measuring}, though \citet{wilson2025no} did not find an association between people's implicit race-status beliefs and their likelihood to select white or non-white candidates in an AI-HITL hiring task. Despite this, IATs are still a tool commonly in workplaces used to both measure and inform people about their implicit beliefs in an effort to change their biased behavior or decision-making \citep{williamson2018unconscious}.

\section{Data and Methods}

\subsection{Independent Variables}

\subsubsection{Occupation Status}
Due to the strength of racial biases in AI resume screening tasks  \cite{wilson2024gender}, \citet{wilson2025no} selected occupations likely to elicit associations between status and racial identity. Prior work has shown that people's perceptions of occupational status is related to the racial composition of its workers \citep{valentino2022constructing} and that people have implicit associations between status and race \citep{melamed2019status, melamed2020measuring}. The set of 16 occupations included in the resume-screening task were eight high status occupations (\textit{sales engineer, construction manager, industrial production manager, nurse practitioner, management analyst, talent agent, computer systems analyst, health services manager}) and eight low status occupations (\textit{agricultural grader, housekeeper, home health aide, textile presser, food preparer, bus person, sales associate, usher}). In each decision scenario, participants were given a short description of the occupation including salary information. Occupation status is represented by the between-subjects factor Status with two levels: High Status and Low Status.

\subsubsection{Candidate Race and Resumes}
\citet{wilson2025no} developed and validated 128 work histories to represent fictitious candidates applying for a given occupation. These candidates had equivalent qualifications and only differed on their implied racial identities. To form a complete resume, each work history was augmented with names and additional interests intended to signal a particular racial identity. The names chosen were highly associated with men who are likely to be one of four race or ethnicities: Asian, Black, Hispanic, or White.\footnote{\citet{wilson2025no} investigated only men of different races to avoid introducing gender confounds into the race-status associations of interest.} Membership in two affinity groups were also included since names are not unambiguously and universally associated with sociodemographic traits \citep{elder2023signaling, gautam2024stop}. These were formed by combining explicit race or ethnicity labels\footnote{Because national and ethnic origin is also highly associated with racial identity \citep{weerts2024unlawful}, membership in both kinds of organizations were included.}, randomly selected leadership positions (President, Vice President, Treasurer, or Secretary), and the name of a randomly selected affinity organization based on those at universities. For Black, Hispanic, and Asian candidates, racial identity was explicitly stated, but white candidates had no explicit race stated to avoid associations with White supremacist movements that could impact quality judgments. Furthermore, in the United States, white identity is often assumed, even when not explicitly labeled, because this is the dominant social group \citep{cheng2023marked}. Examples of names, affinity groups, and a complete resume are available in Appendix \ref{app:a} and \ref{app:b}. Participants saw both White and non-White candidates based on identities signaled in the resume (the within-subjects ``Resume'' factor), but were randomly assigned to a between-subjects ``Group'' factor determining which candidates were compared in the resume-screening task: Asian vs. White, Black vs. White, or Hispanic vs. White.\footnote{While real-world resume screening would not involve only binary identity comparisons, \citet{wilson2025no} chose this formulation in order to align with the race-status IATs which also use binary contrasts.} Additionally, we also encoded the index each candidate resume appeared at within the list of resumes in a given resume-screening scenario as a control variable since positional biases are well-attested in decision-making experiments \citep{bar2015position, joachims2007evaluating, raghubir2006center}. This factor, Position, had the five ordinal levels.

\subsubsection{AI Recommendations}
The six AI recommendations used by \citet{wilson2025no} varied both in magnitude and direction of racial bias: None (no recommendation), Neutral (recommend White and Non-White candidates equally), Congruent/Moderate, Incongruent/Moderate, Congruent/Severe, and Incongruent/Severe. Congruent and Incongruent refer to the preference direction of AI recommendations relative to dominant cultural stereotypes in the United States (i.e., Non-White and White candidates are more likely to be associated with Low Status and High Status occupations, respectively); Moderate and Severe refer to the magnitude of AI bias. In Moderate conditions, candidates were recommended on average 75\% or 25\% of the time (depending on their race and the congruency of the condition), which reflects the actual rate at AI models exhibit racial bias in resume-screening applications \citep{wilson2025no}. In Severe conditions, candidates were recommended 100\% or 0\% of the time (depending on the congruency of the condition). More detailed information about how often candidates were recommended in each condition is available in Appendix \ref{app:c}. These six conditions were designed to represent the most likely types of bias to occur in AI systems, more extreme instances of bias to that could be used to determine impacts on the bounds of human decision-making, and baselines of unbiased systems and systems without AI recommendations. The factor Bias was within-subjects, and each participant saw four levels of the Bias factor, including the None and Neutral levels and both of either the Congruent or Incongruent recommendations. We also recoded the factors Status and Bias into a single binary factor, Recommended (Yes/No), for analyses that did not necessitate more fine-grained delineations.

\subsubsection{Implicit Beliefs and Task Order}
\citet{wilson2025no} assessed participants' implicit associations between status and racial identities using materials from \citet{melamed2019status} and \citet{montgomery2024measuring} in an IAT implemented on Qualtrics with \texttt{iatgen} \citep{carpenter2019survey}. There were three versions of the IAT, each corresponding to one level of Group, and each participant was assigned the same Group for both the resume-screening and IAT tasks. IAT scores were computed using the the algorithm in \citet{greenwald2003understanding}, which gives each participant an effect size score D, where greater positive values mean greater stereotype-congruent associations and smaller negative values mean greater stereotype-incongruent associations. The strength of associations used for IAT scoring corresponds to that of Cohen's D: 0.2 $\leq$ D < 0.5 is considered a small effect, 0.5 $\leq$ D < 0.8 a medium effect, and $\geq$ 0.8 a large effect \citep{cohen2016power}. Participants were not told their D scores after completing the IAT to avoid unblinding them to the purpose of the study. An additional factor, Order, refers to whether or not participants completed IATs before or after the resume-screening task. This was included to test how other aspects of overall system design can influence the effects of biased AI recommendations.

\subsection{Dependent Variables}
The first outcome we were interested in predicting using the variables described in the previous section (Selected) was whether or not a particular candidate's resume was chosen to progress after resume screening (Yes/No). This differed from the target of analysis in \citet{wilson2025no}--while they focused on outcomes at a trial-level (i.e., the racial composition of the entire group of selected candidates), we focus on lower-level outcomes for individual candidates. The second quantity we aimed to model was the amount of time (in seconds) participants spent viewing each resume (Duration). This was recorded by tracking when participants clicked to open and close particular resumes. While this was originally measured with millisecond precision, we rounded each item down to the nearest tenth of a second in order to minimize noise from browser differences. 

\subsection{Participants}
\citet{wilson2025no} gathered responses from 528 participants recruited from Prolific who lived in the United States, spoke English fluently, and did not participate in their earlier related study. Of these, 47.9\% were men; 50.4\% were women, and the remaining 1.7\% were another gender(s). Participants' average age was 39.1 years (SD=11.7). The majority (70.4\%) of participants were White or European alone or in combination with another racial identity; 21.3\% were Black or African alone or in combination with another identity; 7.2\% were Hispanic or Latino/a/x alone or in combination with another identity; 5.0\% were Asian or Asian American alone or in combination with another identity; finally, 1.3\% indicated another race not investigated in this study.\footnote{These proportions do not sum to 100\% because people can belong to more than one group.} Only 30.0\% of participants said they had taken an IAT previously, with the remainder saying they had not or weren't sure.\footnote{\citet{wilson2025no} also asked participants about whether they had hiring experience but found that this did not have an effect on decision-making outcomes in the hiring task. Exploratory analysis shown in Appendix \ref{app:i} also shows no association between people's hiring experience and the time they spent viewing resumes. Therefore, as this study is partially designed to explore the results in \citet{wilson2025no} and we did not have any other hypotheses about participant's hiring experience, we chose not to include it in our main analyses.}

\subsection{Experimental Procedure}

\begin{figure}[!t]
    \centering
    \begin{subfigure}[t]{0.33\textwidth}
        \includegraphics[width=\textwidth]{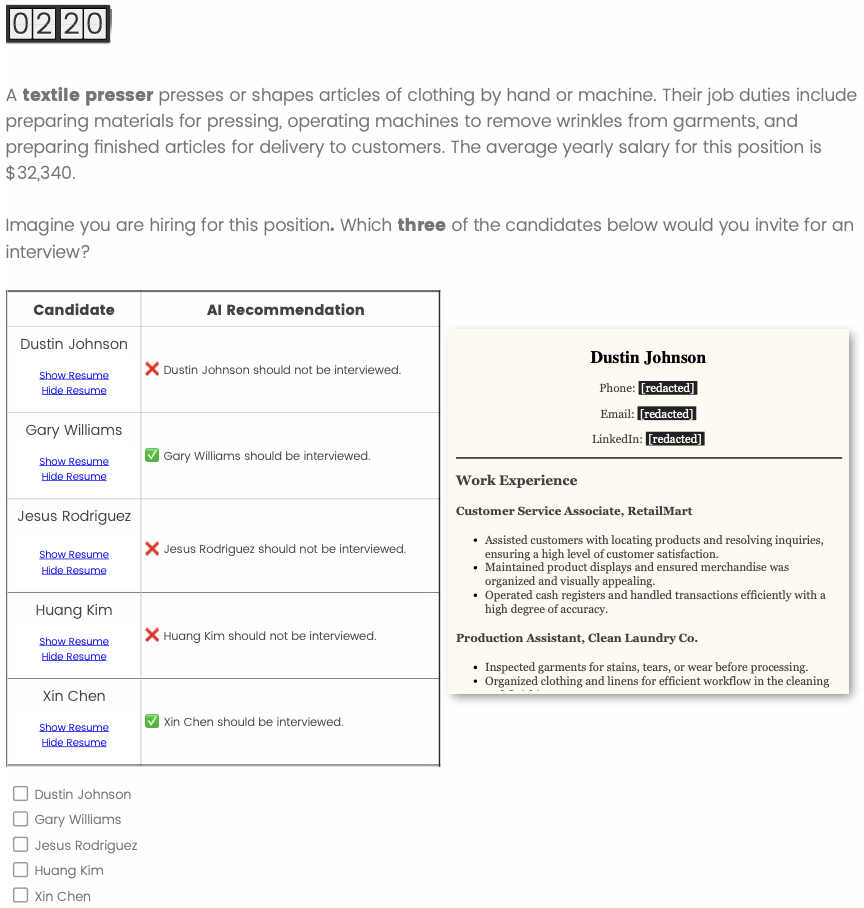}
    \end{subfigure}
    \begin{subfigure}[t]{0.45\textwidth}
        \includegraphics[width=\textwidth]{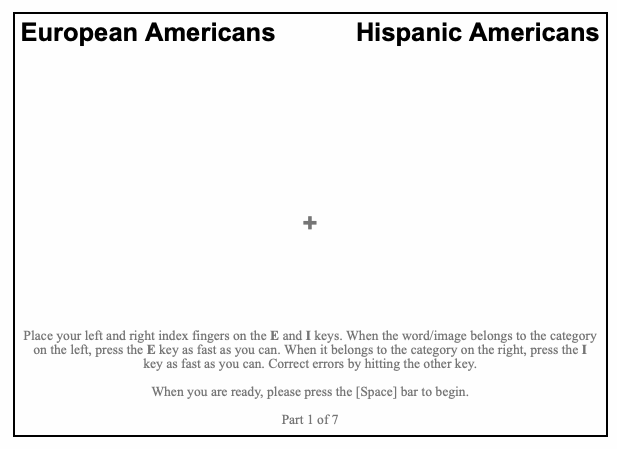}
    \end{subfigure}
    \caption{An example of the interfaces participants saw for the resume-screening task (left) and IAT (right).}
    \label{fig:interface}
\end{figure}

Before beginning the tasks, participants signed a consent form and were randomly assigned to levels of the Group, Status, and Order factors. For Bias, they were randomly assigned a subset of all conditions. Depending on their Order assignment, participants read instructions for either the IAT or the resume-screening task and completed that part of the experiment, followed by the other part. In order to keep participants naive to the true purpose of the study, they were told only that researchers were interested in knowing whether AI recommendations were similar to humans' and if they improved decision-making efficiency. After completing both tasks, participants completed a brief survey and were debriefed to the purpose of the experiment. Finally, each participant was paid a flat rate (based on the minimum wage in Seattle in 2025) for approximately 25 minutes of work.

In the resume-screening task, participants were given a description of an occupation and the names and resumes of five job candidates. There were four qualified resumes, two of which belonged to White candidates and two of which belonged to non-White candidates (either Asian, Black, or Hispanic, depending on the assigned Group condition); the final resume lacked qualifications (as content was written for an occupation different than the one of interest) and was never given a positive AI recommendation. Additionally, this candidate's apparent race was randomly chosen from the identities not in the main comparison. This distractor candidate was included for several reasons: first, having three candidates of different races obscured the true purpose of the experiment and better represented real-world conditions; second, the candidate was unambiguously less qualified and thus served as an attention check if selected. In this case, the trial was excluded from analysis to minimize low-validity data.

Participants had four minutes to review all candidates' resumes and AI recommendations and select three of the five candidates which they thought were most suitable for the given occupation. This amount of time was chosen so that participants spent approximately one minute reviewing each qualified resume in order to align with the time constraints in real-world resume screening \citep{Chan_2024} that might cause decision makers to rely on biased heuristics \citep{kahneman2011thinking}. Once the four minutes had passed, participants could no longer view the resumes and had to submit their choices. 

Participants completed four total trials of the decision task. In the first trial, they saw no AI recommendations, only candidate names and resumes. In the remaining trials, they saw resumes and AI recommendations which were Neutral (recommending exactly one candidate from each comparison race), Congruent/ or Incongruent/Moderate (recommending candidates based on simulated levels of realistic AI racial bias), and Congruent/ or Incongruent/Severe (recommending all candidates from one race and none from the other). The final three trials were always presented in a random order after the first trial, in order to avoid priming participants in scenarios with no AI recommendations.\footnote{We are grateful to a reviewer for suggesting that since participants always see scenarios without AI recommendations first, there may be a confound between trials with and without AI recommendations and scenario ordering. See Appendix \ref{app:h} for further discussion and analysis of this possibility.} An example of the interface participants saw in each trial is shown in Figure \ref{fig:interface}. In each scenario, candidate resumes were presented in a random order. 

In the race-status IAT task adapted from \citet{montgomery2024measuring} and \citet{melamed2019status}, participants sorted words or pictures associated with particular targets (racial identities) or attributes (social statuses) by pressing keys on a keyboard in response to an item appearing on the screen. In the first and second practice blocks, only targets and attributes are sorted, respectively. In blocks three and four, targets and attributes are sorted together. In the remaining blocks, the prior three blocks are repeated with sorting categories appearing in reversed positions on the screen. This task takes approximately five minutes. An example of the interface participants saw for the IAT is in Figure \ref{fig:interface}, and the IAT materials are included in Appendix \ref{app:d}.

\section{Experiments and Results}

From the 1,732 scenarios in the original data from \citet{wilson2025no}, we applied further filtering to remove data points which corresponded to distractor resumes or resumes that had viewing durations longer than 240 seconds, the maximum time allowed for a single scenario. This left 6,417 observations for the analysis procedures and results described below. Complete ANOVA and regression tables for each analysis are available in Appendices \ref{app:e1}-\ref{app:f}.

\newcounter{mybox}
\refstepcounter{mybox}
\begin{tcolorbox}[size=small, title=\text{Box~\themybox: Mixed-effects model specifications},label=box:models,]
\vspace{-\abovedisplayskip}
\begin{align}
\text{\textbf{RQ1}} &\textbf{: } \text{Selected $\sim$ Duration $\times$ Recommendation + Position + (1 | Participant) + (1| Occupation)} \nonumber \\
\text{\textbf{RQ2,3}} &\textbf{: } \text{Duration $\sim$ Resume $\times$ Status $\times$ Bias $\times$ Group $\times$ Order + Position + (1 | Participant) + (1 | Occupation}) \nonumber \\
\text{\textbf{RQ2,4}} &\textbf{: } \text{Duration $\sim$ Resume $\times$ Status $\times$ Bias $\times$ Group $\times$ D + Position + (1 | Participant) + (1 | Occupation}) \nonumber 
\end{align}
\end{tcolorbox}

\subsection{Relationships Between Resume Viewing Duration and Selection (RQ1)} \label{sec:RQ1}

We fitted the binomial logistic model shown in Box \ref{box:models} (estimated using REML and nlminb optimizer) to predict whether a particular candidate was selected based on the interaction between the time participants spent viewing the candidate's resume, whether the candidate was recommended, and random intercepts for Participant and Occupation to account for repeated measures. The model's explanatory power related to the fixed effects alone (marginal R2) is 0.24; conditional R2 could not be computed due to low random effect variance (i.e., there was no correlation between repeated observations). The model's intercept, corresponding to Duration = 0, Recommendation = NA and Position = 0, is at  1.17 (95\% CI [1.08, 1.26], p < .001). To avoid Type I errors, we identify significant factors using a Type III ANOVA omnibus test before conducting post-hoc comparisons to identify differences in factor levels. When warranted, post-hoc tests are computed using a Wald z-distribution approximation with Holm-Bonferroni p-value correction for further Type I error control.

\begin{figure}
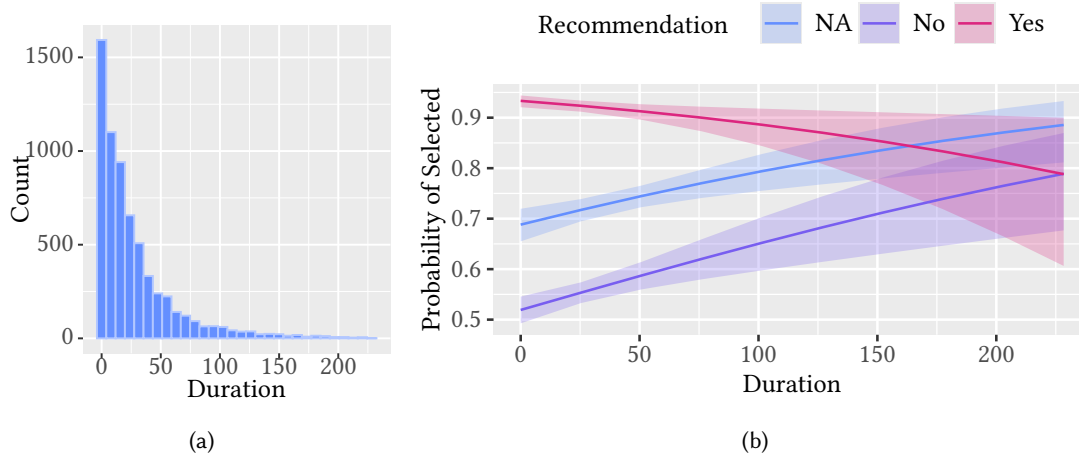

    \centering
    \begin{subfigure}[t]{0.33\textwidth}
        \includesvg[width=\textwidth]{images/p_hist.svg}
        \caption{}
        \label{fig:p_hist}
    \end{subfigure}
    \begin{subfigure}[t]{0.58\textwidth}
        \includesvg[width=\textwidth]{images/p0.svg}
        \caption{}
        \label{fig:p0}
    \end{subfigure}
    \caption{(a) Most resumes were viewed for less than 60 seconds before being selected or not. (b) When candidates are not recommended, viewing resumes for longer results in a significantly higher chance of being selected. When candidates are recommended, longer viewing significantly reduces candidates change of being selected.}
\end{figure}

The distribution of Duration is shown in Figure \ref{fig:p_hist}; on average, participants viewed resumes for 27.2 seconds ($SD=34.51$). For the omnibus main effects, Recommendation was significant ($\chi^2(2) = 549.06$, $p < 0.001$), but Position and Duration were not. The Duration $\times$ Recommendation interaction was also significant ($\chi^2(2) = 21.99$, $p < .001$). This indicates that the influence of Duration differed across Recommendation levels, and post-hoc comparisons of levels of these factors are warranted. \textbf{When no recommendation was given or when a candidate was not recommended by an AI system, increased Duration was significantly associated with higher selection rates} (Recommendation=NA: $\beta = 0.001$, 95\% CI $[0.0005, 0.0017]$, $z = 3.49$, $p < .001$;  Recommendation=No: $\beta = 0.001$, 95\% CI $[0.0007, 0.002]$, $z = 3.81$, $p < .001$). However, \textbf{when candidates were recommended, Duration had a negative association} ($\beta = -0.0004$, 95\% CI $[-0.0007, -0.0001]$, $z = -2.59$, $p < 0.01$). While these results do not establish causal links between Duration and Selection (because scenarios without AI recommendations were always completed first), there are clear trends which warrant further investigation, as shown in Figure \ref{fig:p0}. The largest of these is when a negative AI recommendation is given, such that spending an additional 30 seconds viewing a resume increases that candidate's selection chance by 4\%.

\subsection{Effects of AI Bias and System Design on Viewing Duration (RQ2 and RQ3)} \label{sec:RQ2}

Next, we fitted a model predicting whether interactions between factors related to AI recommendations (i.e., Resume, Status, Bias, Group) and system design (i.e., whether participants were exposed to an IAT before or after completing the AI-HITL resume-screening task) influenced the amount of time spent viewing resumes. The full model specification is given in Box \ref{box:models}. The model's total explanatory power is substantial (conditional R2 = 0.53) and the part related to the fixed effects alone (marginal R2) is 0.12. The model's intercept, corresponding to Resume = Non-White, Status = High-Status, Bias = None, Group = Asian vs. White, Order = Decision/IAT, and Position = First, is at 3.22 (95\% CI [3.16, 3.29], p < .001). For ombibus significance testing, we followed the same procedures to control for Type I error as in Section \ref{sec:RQ1}. For brevity, we report only main effects and the significant interactions which are not entirely contained within another higher-order interaction, with complete omnibus results  in Appendix \ref{app:e}. Because there is no consensus on how to derive standardized effect sizes for mixed effects models \citep{baguley2009standardized, luo2021reporting}, we report mean ratios and their 95\% confidence interval.

There were main effects of Resume ($\chi^2(1) = 4.48$, $p = .034$); Bias ($\chi^2(5) = 305.66$, $p < .001$); and Position ($\chi^2(4) = 289.17$, $p < 0.001$). Resumes were read for significantly less time as their position in the list decreased, with the first resumes being read for approximately 13.19 seconds more than those in the last position ($z=12.95$, $p<.001$, 95\% CI: [10.33, 16.04]). The following interaction effects were significant: Resume $\times$ Status $\times$ Bias ($\chi^2(5) = 13.99$, $p = .016$); Resume $\times$ Status $\times$ Group $\times$ Order($\chi^2(2) = 7.01$, $p = .030$); and Status $\times$ Bias $\times$ Group $\times$ Order ($\chi^2(10) = 21.44$, $p = .018$). To make these higher-order interactions more interpretable while still controlling Type I error rate, we conduct omnibus ANOVAs of lower-order nested interactions and pairwise tests of significant omnibus factors with Holm-Bonferroni p-value corrections to determine which contrasts drive the highest-order interaction \citep{dawson2014moderation}.

\begin{figure}
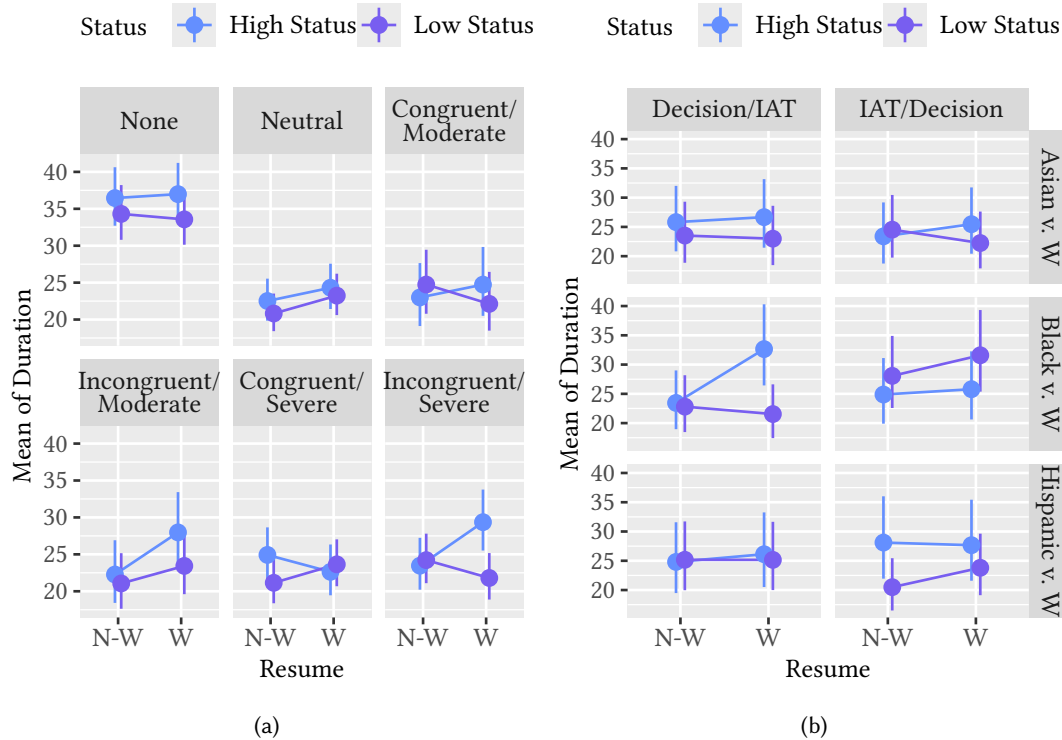

    \centering
    \begin{subfigure}[t]{0.45\textwidth}
        \includesvg[width=\textwidth]{images/p1.svg}
        \caption{}
        \label{fig:p1}
    \end{subfigure}
    \begin{subfigure}[t]{0.45\textwidth}
        \includesvg[width=\textwidth]{images/p2.svg}
        \caption{}
        \label{fig:p2}
    \end{subfigure}
    \caption{(a) These plots show the time spent (y-axis) viewing resumes based on the whether the candidate is perceived as white (W) or non-white (N-W) (x-axis), the kind of occupation in the scenario (color), and the bias of recommendations (panels). Participants spend nearly 13 seconds longer reading resumes without AI recommendations compared to with them. Additionally, in the Incongruent/Severe condition with high status occupations, people view W candidates' resumes (who are not recommended) significantly longer than N-W candidates'. (b) These plots show time spent viewing resumes based on signaled candidate race and occupation type as in (a), but panel columns show whether resume screening preceded or followed the IAT and panel rows specify whether N-W resumes are Asian, Black, or Hispanic. Participants who completed the resume-screening task before the IAT when evaluating Black vs. white resumes spent significantly more viewing the latter for high status jobs. When the IAT was completed first, the time spend viewing white candidates decreases. There are no significant differences when comparing white and Asian or Hispanic candidates.}
\end{figure}

As shown in Figure \ref{fig:p1}, there is a significant Resume $\times$ Status interaction when Bias is Congruent/Severe ($\chi^2(1) = 4.91$, $p = .026$) or Incongruent/Severe ($\chi^2(1) = 9.43$, $p = .002$). Only in the Incongruent/Severe condition are post-hoc pairwise comparisons significant, in which \textbf{resumes of non-white candidates are viewed for significantly less time than those of white candidates when evaluating for a high status occupation} ($z=-2.79$, $p=.005$); this corresponds to participants viewing non-white candidates for approximately 6 seconds less than white candidates (ratio: .799, 95\% CI [.683, .935]). Additionally, there is a main effect of Resume when Bias is Incongruent/Moderate ($\chi^2(1) = 7.11$, $p = .008$), where \textbf{non-white candidates' resumes are viewed for approximately 4 seconds less than white candidates',} a significant difference ($z=-2.18$, $p=.03$, ratio: .846, 95\% CI [.728, .983]). 

As shown in Figure \ref{fig:p2}, there is a significant Resume $\times$ Status $\times$ Order interaction ($\chi^2(1) = 8.65$, $p = .003$) when  comparing Black vs. White resumes, within which Resume $\times$ Status is significant when the resume screening task is completed before the IAT ($\chi^2(1) = 16.93$, $p < .001$). Specifically, pairwise comparisons suggest that \textbf{Black candidates' resumes are viewed for significantly less time than White candidates' when occupations are high status} ($z=-3.87$, $p<.001$, ratio: .718, 95\% CI [.616, .838]). Additionally, when the IAT precedes the resume screening task for Black vs. White candidates, there is a main effect of \textit{Resume} ($\chi^2(1) = 5.44$, $p = .02$), though pairwise comparisons are not significant. There are no significant main or interaction effects when comparing White vs. Asian or Hispanic candidates. 

\begin{figure}
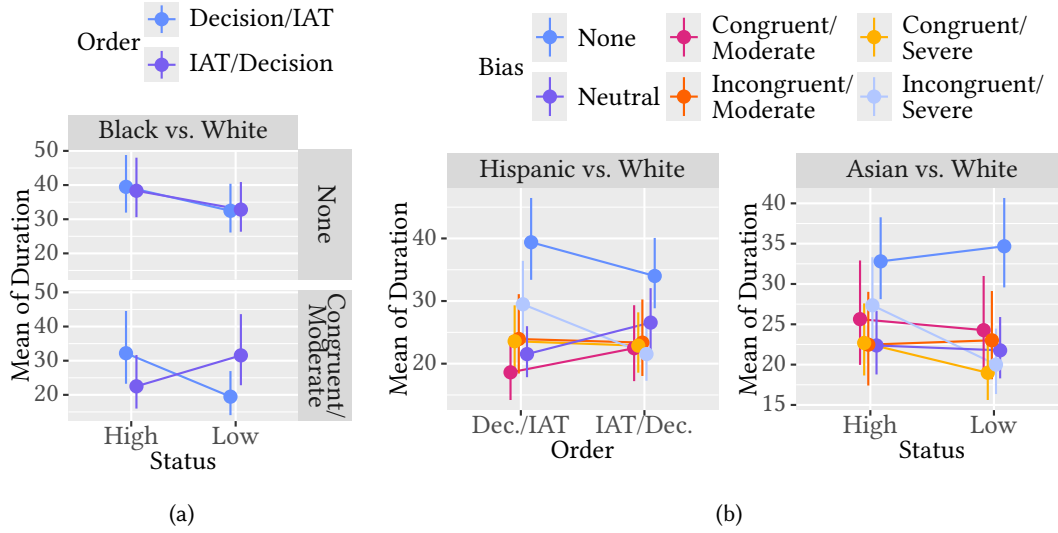

    \centering
    \begin{subfigure}[t]{0.3\textwidth}
        \includesvg[width=\textwidth]{images/p3a.svg}
        \caption{}
        \label{fig:p3a}
    \end{subfigure}
    \begin{subfigure}[t]{0.6\textwidth}
        \includesvg[width=\textwidth]{images/p3bc.svg}
        \caption{}
        \label{fig:p3b}
    \end{subfigure}
    \caption{(a) These plots show the time spent (y-axis) viewing Black vs. white resumes based on the kind of occupation in the scenario (x-axis), the order tasks are completed in (color), and the bias of a subset of recommendations (panel rows). When the resume-screening task is completed first, people spend longer viewing resumes for high status occupations than low status if they view Congruent/Moderate recommendations. (b) These plots show the time spent (y-axis) viewing based on the bias of recommendations (color). For Hispanic vs. white comparisons (left panel), this is further split by the order tasks are completed in (y-axis). In this set of conditions, people spend more time viewing resumes in the Incongruent/Severe condition if they have completed the resume-screening task first. For Asian vs. white comparisons (right panel) split by type of occupation in the scenario (x-axis), resumes were viewed longer when evaluating high status occupations with Incongruent/Severe recommendations.}
\end{figure}

Finally, Figures \ref{fig:p3a} and \ref{fig:p3b} show the significant Status $\times$ Bias $\times$ Group $\times$ Order interaction. For Black vs. White candidates, there was a significant Status $\times$ Bias $\times$ Order interaction ($\chi^2(5) = 13.35$, $p = .020$) within which there were significant Status $\times$ Bias interactions when the resume screening task was completed first ($\chi^2(5) = 13.20$, $p = .021$) or second ($\chi^2(5) = 15.67$, $p = .008$). The only significant pairwise comparison suggests \textbf{people spend significantly longer viewing resumes for high status occupations when recommendations are Congruent/Moderate if the resume-screening task was first} ($z=2.04$, $p=.042$, ratio: 1.653, 95\% CI: [1.043, 2.62]). For Hispanic vs. White candidates, there was a significant Order $\times$ Bias interaction ($\chi^2(5) = 16.01$, $p = .007$) driven by \textbf{resumes being viewed for longer when completing the decision task first in the Incongruent/Severe condition} ($z=2.00$, $p=0.045$, ratio: 1.369, 95\% CI: [1.011, 1.186]). Finally, for Asian vs. White, there is a significant Status $\times$ Bias interaction ($\chi^2(5) = 15.67$, $p = .008$) driven by \textbf{resumes being viewed for longer with high status occupations and Incongruent/Severe recommendations} ($z=2.14$, $p=.032$, ratio: 1.368, 95\% CI: [1.032, 1.81]). 

For all groups, there was main effect of Bias ($\chi^2(5) \geq 106.00$, $p < .001$), also suggesting people may spend more time viewing resumes when they were not given recommendations ($z \geq 8.78$, $p<.001$). A similar result is also found for the main effect of Bias--resumes are viewed for significantly longer when no AI recommendations are given (Bias = None), compared to any other level of Bias and regardless of the type of resume or occupation status ($z \geq 9.15$, $p < .001$). In the most extreme case, resumes were viewed nearly 13 seconds longer when no recommendations were given compared to Neutral recommendations (ratio: 1.556, 95\% CI [1.425, 1.70]), though this effect could also be partially attributed to the order of scenarios, as discussion in Appendix \ref{app:i}.

\subsection{Effects of AI Bias and Implicit Beliefs on Viewing Duration (RQ2 and RQ4)}

\begin{figure}
    \includesvg[width=\textwidth]{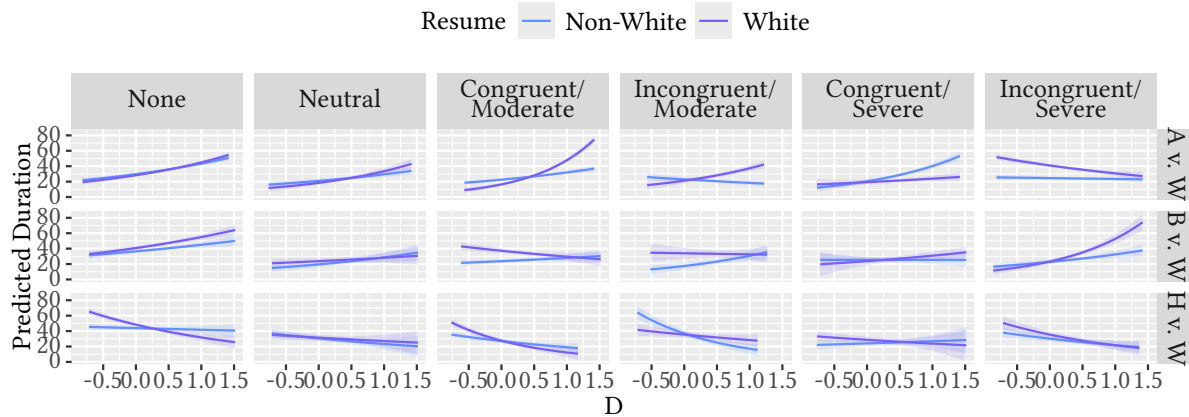}
    \caption{This plot shows the change in time spent (y-axis) viewing resumes as participants' IAT score changes (x-axis) based on the bias of recommendations (panel columns) and whether candidates are Asian, Black, Hispanic, or white (panel rows and color). When comparing Asian vs. white (A v. W) candidates, more stereotypical associations between white and high status (more positive values of D) correspond to spending longer viewing white resumes. For Black vs. white (B v. W) comparisons, higher D values are associated with longer viewing times regardless of candidate race. For Hispanic vs. white (H v. W) comparisons, D has a negative effect on viewing time when recommendations are Congruent/Moderate.}
    \label{fig:p5}
\end{figure}

Lastly, to determine how interactions between AI recommendations and individuals' implicit associations influenced the amount of time spent viewing resumes, we used the same general linear mixed modeling procedure as in Section \ref{sec:RQ2}, except investigating D in place of Order (model specification is given in Box \ref{box:models}). The model's explanatory power is substantial (conditional R2 = 0.53) and the part related to the fixed effects (marginal R2) is 0.13. The model's intercept, corresponding to Resume = Non-White, Status = High Status, Bias = None, Group = Asian vs. White, Order = Decision/IAT, and Position = First, is at 3.18 (95\% CI [3.10, 3.26], p < .001). For omnibus significance testing, we also follow the procedures described in Section \ref{sec:RQ1} for controlling Type I error rates, with full results in Appendix \ref{app:f}. Effect sizes for effects involving D are reported using model coefficients rather than mean ratios as D is continuously valued.

There was a main effect of Position ($\chi^2(4) = 294.05$, $p < 0.001$), as first resumes were read 13.38 seconds more than the last ones ($z=13.05$, $p<.001$, 95\% CI: [10.50, 16.26]). Bias was also significant as a main effect ($\chi^2(5) = 152.23$, $p < .001$) and in two interactions: Status $\times$ Bias $\times$ Group ($\chi^2(10) = 31.228$, $p < .001$) and Resume $\times$ Bias $\times$ Group $\times$ D ($\chi^2(10) = 19.27$, $p = .037$). As in Section \ref{sec:RQ2}, these higher-order interactions are difficult to interpret in isolation, so we decompose them to identify significant two-way differences. 

As shown in Figure \ref{fig:p5}, although there was a significant four-way Resume $\times$ Bias $\times$ Group $\times$ D interaction, there were no significant three-way interactions at any level of \textit{Group}. However, for Black vs. White, there were significant main effects of Resume ($\chi^2(1) = 10.35$, $p =.001$), Bias ($\chi^2(5) = 77.38$, $p < .001$), and D ($\chi^2(1) = 4.69$, $p = .030$). \textbf{Post-hoc pairwise comparisons suggest that resumes of Black candidates are read for significantly less time than white candidates'} ($z=-2.85$, $p=.004$, ratio: .891, 95\% CI: [.824, .964]); \textbf{subjects spend more time reading resumes when there were no recommendations; and people with stronger stereotypical implicit beliefs (more positive \textit{D} scores) spend longer reading resumes} ($z=2.44$, $p=.015$, $\beta = .27$, 95\% CI: [.05, .48]). 

For Asian vs. White candidates, there were significant effects of Bias ($\chi^2(5) = 77.38$, $p < .001$) and D ($\chi^2(1) = 4.96$, $p = .026$), Resume $\times$ Bias ($\chi^2(5) = 13.73$, $p =.017$) and Resume $\times$ D ($\chi^2(1) = 5.62$, $p = .018$). Pairwise comparisons suggest that \textbf{when recommendations were Congruent/Moderate, Asian candidates' resumes were viewed for significantly longer than white candidates'} ($z=2.28$, $p=.023$, ratio: 1.368, 95\% CI: [1.052, 1.778]). Additionally, the significant Status $\times$ Bias interaction ($\chi^2(5) = 14.46$, $p =.013$) suggests that \textbf{Asian and white resumes were viewed for significantly longer when evaluating high status occupations with Incongruent/Severe recommendations} ($z=2.16$, $p=.031$, ratio: 1.176, 95\% CI: [1.035, 1.82]). Furthermore, \textbf{D positively influenced the resume viewing duration for White candidates} ($z=2.72$, $p=.007$, $\beta = .33$, 95\% CI: [0.09, 0.57]). 

For Hispanic vs. White comparisons, there was a significant main effect of Bias ($\chi^2(5) = 56.36$, $p < .001$) and Bias $\times$ D interaction ($\chi^2(1) = 19.74$, $p = .001$), which was due to a \textbf{negative association between D and viewing duration when recommendations were Congruent/Moderate} ($z=-2.33$, $p=.020$, $\beta = -.50$, 95\% CI: [-0.92, -0.08]). Finally, as above, post-hoc tests suggest that \textbf{the effects of Bias for each group may be driven by differences between conditions with and without recommendations} ($z \geq 4.20$, $p<.001$).

\section{Discussion}

\subsection{Procedural vs. Distributive Fairness}

In the first analysis, we identified a significant relationship between AI recommendations and the time spent viewing resumes that predicted whether they would be selected. Namely, we observed that when a candidate was not recommended, spending more time reviewing their resume led to a higher chance of being selected while the reverse was true for candidates that were recommended. This suggests that AI recommendations may serve as heuristic cues that guide people to quick and easy decisions, though further experiments modeling additional aspects of decision making are needed to establish their strength. However, the more that people elaborated over a resume, the more likely they were to contradict AI recommendations, suggesting the importance of humans-in-the-loop that are more deeply and thoughtfully engaging with these decisions. Since all candidates were equally qualified in this experiment and coming to unbiased decisions often required contradicting AI recommendations, designing AI systems which encourage people to engage more analytical, rational types of thinking (such as that used by \citet{rastogi2022deciding} to mitigate anchoring bias) could be a potential route to increase the effectiveness of human oversight in high-stakes AI-HITL tasks. 

The existence of a relationship between the time spent viewing resumes and selection of resumes also has implications for different kinds of fairness in hiring. \citet{fabris2025fairness} catalogue a number of fairness metrics that have been applied to AI systems used for hiring tasks, the majority of which operate solely over the outcomes and are therefore measures of distributive fairness or justice. By measuring the duration of time people spend viewing resumes, we show that even within the same task completed by the same participants, procedural fairness evaluations may diverge from distributive fairness evaluations, even though they are inherently related \citep{barocas-hardt-narayanan, trautmann2023procedural}. For example, while we found that differences in viewing durations tend to be greatest for high status occupations and negligible for low status occupations, \citet{wilson2025no} showed that outcomes were uniformly affected by AI recommendations regardless of occupation status. 

These fairness evaluation differences also suggest a need to use a variety of methods to investigate employment discrimination, since many laws and regulations can be understood as prohibiting either distributive or procedural unfairness \citep{selmi2023disparate}. As distributive outcomes are both easier to assess and for employees to recognize due to the general opacity of hiring processes, procedural unfairness might be especially difficult to prevent or repair. Therefore, we believe that as AI becomes more intertwined into hiring processes, there will be a greater necessity for anti-discrimination employment laws to be enforced proactively instead of retroactively and on a wide scale instead of on a case-by-case basis. For example, in the United States, the Equal Employment Opportunity Commission could use its powers to prevent systemic discrimination to engage in information gathering even in settings where discrimination has not been reported previously \citep{kim2015addressing}. Additionally, we suggest that future evaluations of AI-HITL systems consider both procedural and distributive fairness not only for compliance, but also to maximize people's trust in systems \citep{valcke2020procedural, choung2025fairness} and their satisfaction with outcomes \citep{bedemariam2023roles, greenberg2012promote, brockner2013and}.

\subsection{Fast vs. Slow Resume Screening}

Crucially, people spent considerably less time viewing resumes when paired with AI recommendations as compared to without AI recommendations, with one possible explanation being that AI recommendations in hiring tasks reduce people's use of System 2 analytical thinking, which is critical to prevent people's implicit biases or AI biases from unintentionally influencing behaviors and outcomes. \citet{wilson2025no} found that decisions made without AI recommendations were unbiased; bias only emerged when interacting with biased recommendations. While efficiency and time savings are often stated benefits of AI resume-screening systems \citep{hirevue}, these results show that this may come at the expense of high-quality decisions that balance candidate qualifications with other important factors such as non-discrimination or company fit if models are not appropriately trained on these dimensions prior to human interaction. 

The instances where people did show evidence of engaging relatively more analytical reasoning in the presence of AI recommendations benefited racial groups unequally. When AI exclusively recommended non-white candidates for high status occupations, participants spent more time evaluating white candidates' resumes (which were not recommended) than non-white candidates' resumes. This pattern of results may be consistent with broader theoretical perspectives suggesting people are more likely to engage in more effortful elaboration when decisions are seen as important or when information contradicts people's existing views \citep{oxford_dual, petty1986elaboration, ditto1992motivated, chaiken1996beyond}. Accordingly, AI recommendations that were incongruent with white-high status stereotypes may be threatening and motivate people to defend the status quo, which could explain more time spent scrutinizing these resumes \citep{WilkinsClaraL.2014RPaT}. Notably, the same deliberation was not applied to conditions in which white/non-white candidates were/were not recommended for low status jobs, which also contradict typical associations. This is in-line with social dominance theories, which postulate that downward mobility is preferable to upward mobility because this maintains existing hierarchies rather than inverting them \citep{oxford_group}. 

We observed other effects that can also be explained by System 2 reasoning being activated when decisions are thought to be particularly important, also aligning with work showing that having high intrinsic \citep{buccinca2021trust} or extrinsic \citep{eisbach2023optimizing} motivation can enhance AI-HITL decision making in other tasks. For Asian vs. white candidates, resumes were viewed longer when evaluating high status occupations with Incongruent/Severe recommendations. In contrast to the effect for Black vs. white candidates, there were no significant differences in durations for Asian vs. white resumes, likely because neither of these groups are stereotypically categorized as low status \citep{fiske2018model}. Similarly, when the resume-screening task is completed first and comparing Black vs. white candidates, people spend longer viewing resumes for high status occupations than low status if they view Congruent/Moderate recommendations. Again no differences emerged between candidates of different races because recommendations aligned with common societal stereotypes and participants did not need to reanalyze their initial impressions. Overall, this suggests that there is no one-size-fits-all solution to encourage analytical thinking because ``importance" in decision-making is relative: it emerges for different reasons in different contexts, and system users and developers should be aware of how those factors could encourage imbalanced heuristic/analytical reasoning for different targets of decision making.

\subsection{Reducing Bias through System Design}

An intervention that can potentially change decision-making behavior is completing an IAT before the resume-screening task. We observed that participants only spent an equal time evaluating Black and white candidates for high status jobs if they completed the resume-screening task second (as a result of spending less time evaluating white candidates). Additionally, when comparing Hispanic and white candidates, people spent less time viewing resumes in the Incongruent/Severe condition if they completed the IAT first. Since overall viewing time did not increase, it is unlikely that these differences are a result of less heuristic reasoning. Rather, the explicit naming of race and status words in the IAT may activate a goal to avoid race-based employment discrimination rather than stereotypical race-status associations, which can change the results of System 1 heuristic processing without increasing System 2 analytical reasoning \citep{moskowitz2011egalitarian}. This hypothesis is supported by similar studies in politics, showing that participants who are primed with explicit racial cues are less likely to act in accordance with racial stereotypes than those primed with implicit racial cues \citep{RacialPrimingwithImplicitandExplicitMessages}. Overall, while these results show how system design choices can positively impact human decision-making dynamics, they may not do so by increasing System 2 reasoning. Designers of interventions and mitigation strategies may benefit from considering how a particular method interacts with different decision-making strategies and if it is compatible with their overall goals.

\subsection{Implicit Beliefs}

Finally, people's implicit beliefs also played a role in their time spent viewing resumes, but effects were not consistent across groups. For Black vs. white candidates, more stereotypical beliefs about race and status predicted longer durations viewing resumes, potentially because people with more stereotypical beliefs were also more motivated to carefully consider decisions related to race and status.\footnote{D can also be correlated with demographic traits such as age \citep{nosek2007pervasiveness}, which may also contribute to this process. We discuss this further in Appendix \ref{app:g}.} A similar positive effect of D on viewing durations of white resumes was also found when comparing Asian vs. white candidates, but for Hispanic vs. white comparisons, D had a negative effect on viewing time when recommendations were Congruent/Moderate. The diversity of these results underscores the differing effects that individual traits and settings can have on AI-HITL decision-making processes--it is rarely possible to devise solutions that can satisfy every user's needs, and what works for one application may have the opposite effect in another. By researching decision-making processes in addition to outcomes, we aim to shed light on these differences that can lead to more effective system development, evaluation, use, and oversight.

\subsection{Limitations and Future Work}

This work represents an initial step into a larger research agenda investigating how AI can influence decision-making processes in AI-HITL tasks and designing interactions which allow people to effectively oversee AI decisions. As with many controlled experiments, we simplified the complexity of hiring scenarios encountered in the real world for this analysis. For example, recruiters or hiring managers may have different incentives or willingness to incorporate AI into their workflows, which we did not investigate. Additionally, although participants' self-reported hiring experience was not a significant predictor of decision-making time in this experiment, it's possible that quantifying or characterizing this variable in a more fine-grained way could reveal important differences. Finally, we were limited to analyzing timing, though other artifacts of decision-making processes may also be informative. Future work should continue to evaluate AI systems in the context of their interactions with humans with accurate real-world models, and model developers and users should incorporate these evaluations into the design and implementation of their AI-HITL systems.

\section{Conclusion}

Motivated by the findings of \citet{wilson2025no}, we reanalyze the data from their AI-HITL resume-screening experiment in order to characterize the decision-making dynamics that led people to make racially biased vs. unbiased selections when using (biased) AI recommendations. We found that in most cases, AI recommendations may act as heuristics for decision making and people need to engage deeper, more analytical methods of thinking in order to contradict them. This analytical thinking was most apparent in conditions where white candidates were not recommended for high status jobs, in line with broader psychological theories about motivation and social dominance. By completing an IAT before the resume-screening task, participants were able to lessen differences in time spent evaluating resumes of different races, but this did not increase the time spent evaluating resumes overall. Finally, we identified some tentative relationships between people's implicit beliefs about race and status and the time they spent viewing resumes, but additional research is needed to discern the mechanisms underlying these relationships. All together, these results suggest developing AI-HITL systems that could encourage more analytical and less heuristic methods of reasoning, enabling effective oversight when they are used for high-stakes tasks like resume screening. 

\section*{Generative AI Usage Statement}
The authors of this paper did not use generative AI tools during the preparation of this manuscript.

\begin{acks} 

We are grateful to the anonymous reviewers for their constructive feedback. This work was supported in part by the U.S. National Institute of Standards and Technology (NIST) Award 60NANB23D194; Schmidt Sciences Award on AI and Advanced Computing, through the Science of Trustworthy AI program; the Robert L. McDevitt, K.S.G., K.C.H.S. and Catherine H. McDevitt L.C.H.S. Chair in Computer Science at Georgetown University; and the University of Washington Tech Policy Lab. Any opinions, findings, and conclusions or recommendations expressed in this material are those of the authors and do not necessarily reflect those of supporters. We would also like to express our gratitude to students in the Information School at the University of Washington and friends and family of the authors for their help in piloting the experiments. Finally, we extend the biggest note of appreciation to Prof. Yoshi Kohno, whose guidance and feedback strengthened this work immeasurably and who was a gracious and generous mentor to all of the paper's authors. 

\end{acks}

\bibliographystyle{ACM-Reference-Format}
\bibliography{custom}

\appendix

\section{Demographic Signals on Resumes} \label{app:a}
\begin{table}[h!]
\centering
\begin{tabular}{@{}lll@{}}
\toprule
\textbf{Group}& \textbf{Feature}                                                                 & \textbf{Values}                                                                                                                                                                                                      \\ \midrule
Asian        & First Name                                                             & Hong, Huang, Xin, Yong                                                                                                                                                                                      \\
             & Last Name                                                              & Chen, Kim, Nguyen, Tran                                                                                                                                                                                     \\ \midrule
Black        & First Name                                                             & Jamal, Leroy, Mohammad,  Lamar                                                                                                                                                                               \\
             & Last Name                                                              & \begin{tabular}[c]{@{}l@{}}Jefferson, Johnson, Washington, \\ Williams \end{tabular}                                                                                                                         \\ \midrule
Hispanic     & First Name                                                             & Alejandro, Jesus, Pablo, Santiago                                                                                                                                                                        \\
             & Last Name                                                              & \begin{tabular}[c]{@{}l@{}} Hernandez, Lopez, Martinez, \\Rodriguez\end{tabular}                                                                                                                            \\ \midrule
White        & First Name                                                             & Brent, Dustin, Gary, Todd                                                                                                                                                                                  \\
             & Last Name                                                              & 
             Johnson, O’Brien, Miller, Williams                                                                                                                                                                          \\ \midrule
\textit{All} & \begin{tabular}[c]{@{}l@{}}Racial \\ Affinity \\ Org.\end{tabular} & \begin{tabular}[c]{@{}l@{}}\{Asian, Black, Hispanic, $\emptyset$\} Student \\ Action Association, \_\_\_ Student \\Association, \_\_\_ Student  Leadership \\Coalition, \_\_\_ Student Union\end{tabular}         \\
             & \begin{tabular}[c]{@{}l@{}}Ethnic \\ Affinity \\ Org.\end{tabular} & \begin{tabular}[c]{@{}l@{}}\{Chinese American, Haitian \\ American, Mexican American, \\ English American\} Association, \\ \_\_\_ Heritage Club, \_\_\_ Society, \\ \_\_\_ Youth  Organization \end{tabular} \\
\midrule
\end{tabular}
\caption{Features used on resumes to signal candidates' racial identity.}
\label{features_table}
\end{table}
\clearpage

\section{Example Resume} \label{app:b}
\begin{figure}[h!]
    \centering
    \includegraphics[width=0.6\textwidth]{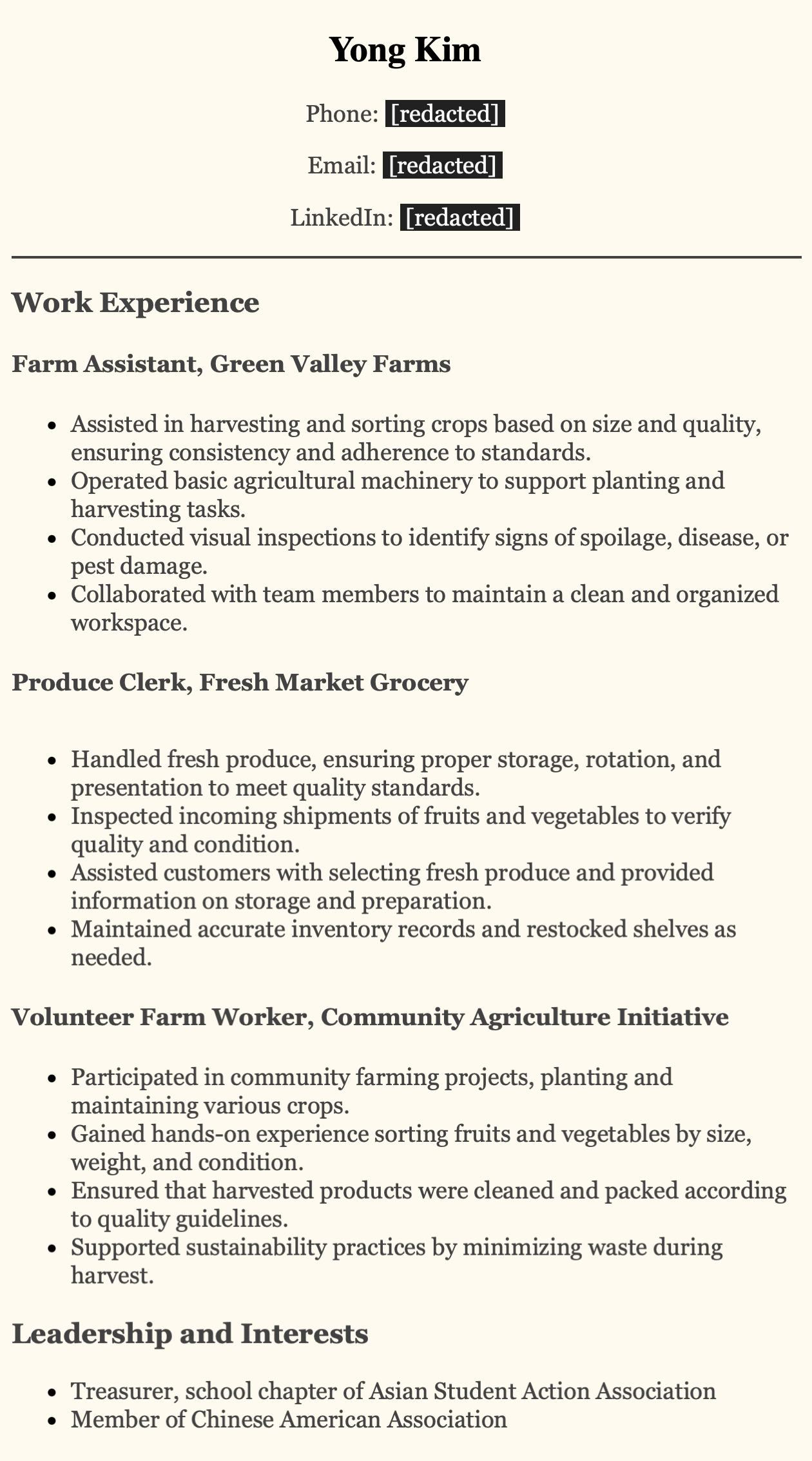}
    \caption{An example of the resumes participants saw for candidates in the resume-screening task.}
    \label{fig:appendix_resumes}
\end{figure}
\clearpage

\section{AI Recommendation Conditions} \label{app:c}
\begin{table}[h!]
\centering
\begin{tabular}{@{}llccc@{}}
\toprule
\textbf{AI Rec.}                                                       & \textbf{Job Status} & \textbf{\begin{tabular}[c]{@{}l@{}}White vs. \\Black\end{tabular}} & \textbf{\begin{tabular}[c]{@{}l@{}}White vs. \\Asian\end{tabular}} & \textbf{\begin{tabular}[c]{@{}l@{}}White vs. \\Hispanic \end{tabular}} \\ \midrule
\multirow{2}{*}{None}                                                  & High                & N/A                                                                 & N/A                                                                 & N/A                                                                    \\
                                                                       & Low                 & N/A                                                                 & N/A                                                                 & N/A                                                                    \\ \midrule
\multirow{2}{*}{Neutral}                                               & High                & .500                                                                  & .500                                                                  & .500                                                                     \\
                                                                       & Low                 & .500                                                                  & .500                                                                  & .500                                                                     \\ \midrule
\multirow{2}{*}{\begin{tabular}[c]{@{}l@{}}Cong/\\ Mod\end{tabular}}   & High                & \begin{tabular}[c]{@{}c@{}}.835\\ (.690 / .980)\end{tabular}          & \begin{tabular}[c]{@{}c@{}}.765\\ (.680 / .850)\end{tabular}          & \begin{tabular}[c]{@{}c@{}}.610\\ (.470 / .750)\end{tabular}              \\
                                                                       & Low                 & \begin{tabular}[c]{@{}c@{}}.830\\ (.870 / .790)\end{tabular}           & \begin{tabular}[c]{@{}c@{}}.695\\ (.680 / .710)\end{tabular}          & \begin{tabular}[c]{@{}c@{}}.770\\ (.880 / .660)\end{tabular}              \\ \midrule
\multirow{2}{*}{\begin{tabular}[c]{@{}l@{}}Cong/\\ Sev\end{tabular}}   & High                & 1.000                                                                   & 1.000                                                                   & 1.000                                                                      \\
                                                                       & Low                 & 0.000                                                                   & 0.000                                                                   & 0.000                                                                      \\ \midrule
\multirow{2}{*}{\begin{tabular}[c]{@{}l@{}}Incong/\\ Mod\end{tabular}} & High                & \begin{tabular}[c]{@{}c@{}}.165\\ (.390 / .020)\end{tabular}          & \begin{tabular}[c]{@{}c@{}}.235\\ (.320 / .150)\end{tabular}          & \begin{tabular}[c]{@{}c@{}}.390\\ (.530 / .250)\end{tabular}              \\
                                                                       & Low                 & \begin{tabular}[c]{@{}c@{}}.170\\ (.130 / .210)\end{tabular}           & \begin{tabular}[c]{@{}c@{}}.305\\ (.320 / .290)\end{tabular}          & \begin{tabular}[c]{@{}c@{}}.230\\ (.120 / .340)\end{tabular}              \\ \midrule
\multirow{2}{*}{\begin{tabular}[c]{@{}l@{}}Incong/\\ Sev\end{tabular}} & High                & 0.000                                                                   & 0.000                                                                   & 0.000                                                                      \\
                                                                       & Low                 & 1.000                                                                   & 1.000                                                                   & 1.000                                                                      \\ \bottomrule
\end{tabular}
\caption{Proportion of simulated AI recommendations that favor White candidates in various combinations of Group, Status, and magnitude and direction of AI recommendation Bias. For Moderate bias conditions, two values are given for jobs with worker demographics that approximate the overall US population vs. those that do not. The results in this paper are presented in terms of the average of these values.}
\label{tab:recs}
\end{table}

\section{IAT Materials} \label{app:d}

\begin{figure}[h!]
    \centering
    \includesvg[width=0.6\textwidth]{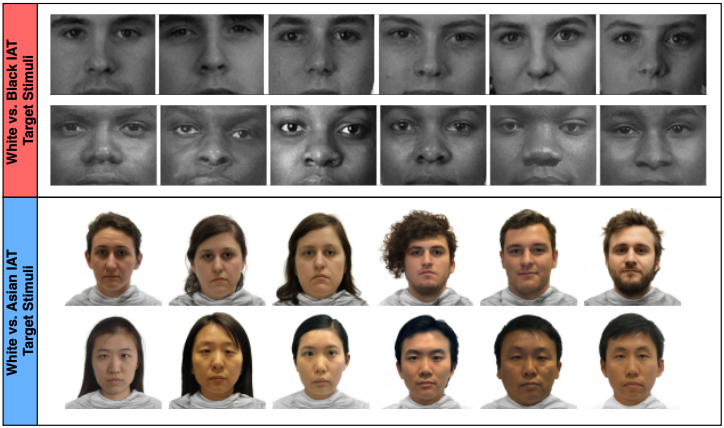}
    \caption{Pictures used to represent racial groups in white vs. Black and white vs. Asian IATs.}
    \label{fig:appendix_iat}
\end{figure}

\subsection{Status Attributes}
Words representing high vs. low status words came from \citet{montgomery2024measuring}, who developed a Status-Gender IAT by testing a variety of words representing status and selecting those which were most associated with implicit categorization. The best set of low-status and high-status words was \textit{(in)capable}, \textit{(in)competent}, \textit{(un)able}, \textit{(un)worthy}, and \textit{(un)skilled}; all of these were used in our study.

\subsection{Race Targets}
The stimuli used for race targets was the same as used in IATs hosted on Project Implicit.\footnote{https://implicit.harvard.edu/implicit/} For white vs. Black and white vs. Asian targets, these were images of faces (shown in Figure \ref{fig:IAT_pics}) which were also used in Race-Status IATs by \citet{melamed2019status}. For white vs. Hispanic targets, these were last names \textit{\{Jones, Davis, Thompson, Smith, Kelly, McDonald\}} for white targets and \textit{\{Torres, Flores, Rivera, Pérez, Sánchez, Ramos\}} for Hispanic targets. Following other IATs on Project Implicit, conceptual terms for racial groups were 
\textit{\{European Americans, African Americans, Asian Americans, Hispanic Americans\}} rather than \textit{\{white, Black, Asian, Hispanic\}} as has been used throughout this paper.

\section{RQ1 Regression Table} \label{app:e1}

This section provides regression tables for the fit logistic mixed effects models following the guidelines given in \citet{meteyard2020best}. In order to conserve space, factor levels are represented by numeric values in the regression tables. Table \ref{tab:reg_keys} provides a key translating factor level names and numeric values, as well as specifying reference levels.

\begin{table}[]
\begin{tabular}{@{}lcccccc@{}}
\toprule
\multicolumn{1}{c}{\textbf{Factor}} & \multicolumn{6}{c}{\textbf{Levels}} \\ 
 & Reference & 1 & 2 & 3 & 4 & 5 \\ \midrule
Bias & None & Neutral & \begin{tabular}[c]{@{}l@{}}Congruent/\\ Moderate\end{tabular} & \begin{tabular}[c]{@{}l@{}}Incongruent/\\ Moderate\end{tabular} & \begin{tabular}[c]{@{}l@{}}Congruent/\\ Severe\end{tabular} & \begin{tabular}[c]{@{}l@{}}Incongruent/\\ Severe\end{tabular} \\
Position & First & Second & Third & Fourth & Fifth &  \\
Group & Asian vs. White & Black vs. White & Hispanic vs. White &  &  &  \\
Resume & Non-White & White &  &  &  &  \\
Status & High & Low &  &  &  &  \\
Order & Decision/IAT & IAT/Decision &  &  &  &  \\ \bottomrule
\end{tabular}
\caption{Key translating factor level names to numeric values used in regression output tables.}
\label{tab:reg_keys}
\end{table}

\begin{longtable}{rcccc}

\toprule

\multicolumn{5}{c}{\textbf{Model Fit}} \\ 
\midrule
\multicolumn{5}{l}{\begin{tabular}[c]{@{}l@{}}Formula: Selected $\sim$Duration $\times$ Recommendation +  Position + (1 | Participant) + (1 | Occupation)\end{tabular}}  \\
\multicolumn{5}{l}{\begin{tabular}[c]{@{}l@{}}Standardized parameters were obtained by fitting the model on a standardized version of the dataset. \\ 95\% confidence intervals and p-values were computed using a Wald z-distribution approximation.\end{tabular}}  \\ 
\bottomrule

\multicolumn{1}{l}{} &  &  &  &   \\ 

\toprule
\multicolumn{5}{c}{\textbf{Random Effects}} \\
\multicolumn{1}{l}{} &  &  & Variance & SD  \\ 
\midrule
\multicolumn{1}{l}{Participant} &  &  & 7.498e-10 & 2.738e-05  \\
\multicolumn{1}{l}{Occupation} &  &  & 5.401e-11 & 7.349e-06 \\
\bottomrule

\multicolumn{1}{l}{} &  &  &  &   \\ 

\toprule
\multicolumn{5}{c}{\textbf{Fixed Effects}} \\
 & Estimate & Std. Error & z value & Pr($>$\$|$z$|\$) \\ 
\midrule
\endfirsthead

{{\tablename\ \thetable{} -- continued from previous page}} \\
\toprule
\multicolumn{5}{c}{\textbf{Fixed Effects}}\\
 & Estimate & Std. Error & z value & Pr($>$\$|$z$|\$)\\ 
\midrule
\endhead

\midrule
\multicolumn{3}{r}{Continued on next page} \\
\midrule
\endfoot

\bottomrule
\caption{Regression fit information for RQ1.}
\endlastfoot

(Intercept) & 1.17 & 0.04 & 26.12 & 0.00 \\ 
  Duration & 0.00 & 0.00 & 1.69 & 0.09 \\ 
  Recommendation1 & -0.38 & 0.06 & -6.02 & 0.00 \\ 
  Recommendation2 & -1.09 & 0.05 & -20.08 & 0.00 \\ 
  Position1 & 0.10 & 0.06 & 1.62 & 0.11 \\ 
  Position2 & -0.01 & 0.06 & -0.18 & 0.86 \\ 
  Position3 & -0.01 & 0.06 & -0.14 & 0.89 \\ 
  Position4 & -0.12 & 0.06 & -2.01 & 0.04 \\ 
  Duration:Recommendation1 & 0.00 & 0.00 & 2.84 & 0.00 \\ 
  Duration:Recommendation2 & 0.00 & 0.00 & 2.87 & 0.00 \\ 

\end{longtable}
\section{RQ2 and RQ3 Omnibus ANOVA and Regression Table} \label{app:e}
\begin{table}[h!]
\begin{tabular}{lrrrllrrr}
\cline{1-4} \cline{6-9}
 & Chisq & Df & Pr($>$Chisq) &  &  & Chisq & Df & Pr($>$Chisq) \\ \cline{1-4} \cline{6-9} 
(Intercept) & 9744.49 & 1 & 0.0000 &  & Resume:Status:Bias & 13.99 & 5 & 0.0157 \\
Resume & 4.48 & 1 & 0.0344 &  & Resume:Status:Group & 3.47 & 2 & 0.1767 \\
Status & 1.44 & 1 & 0.2309 &  & Resume:Bias:Group & 9.17 & 10 & 0.5161 \\
Bias & 305.66 & 5 & 0.0000 &  & Status:Bias:Group & 23.33 & 10 & 0.0096 \\
Group & 0.99 & 2 & 0.6107 &  & Resume:Status:Order & 3.81 & 1 & 0.0510 \\
Order & 0.08 & 1 & 0.7831 &  & Resume:Bias:Order & 7.34 & 5 & 0.1964 \\
Position & 289.17 & 4 & 0.0000 &  & Status:Bias:Order & 15.09 & 5 & 0.0100 \\
Resume:Status & 2.23 & 1 & 0.1355 &  & Resume:Group:Order & 0.73 & 2 & 0.6947 \\
Resume:Bias & 6.87 & 5 & 0.2301 &  & Status:Group:Order & 4.10 & 2 & 0.1288 \\
Status:Bias & 1.20 & 5 & 0.9446 &  & Bias:Group:Order & 14.51 & 10 & 0.1511 \\
Resume:Group & 3.65 & 2 & 0.1613 &  & Resume:Status:Bias:Group & 12.08 & 10 & 0.2797 \\
Status:Group & 0.39 & 2 & 0.8237 &  & Resume:Status:Bias:Order & 7.25 & 5 & 0.2026 \\
Bias:Group & 13.85 & 10 & 0.1801 &  & Resume:Status:Group:Order & 7.01 & 2 & 0.0301 \\
Resume:Order & 0.04 & 1 & 0.8356 &  & Resume:Bias:Group:Order & 11.30 & 10 & 0.3343 \\
Status:Order & 0.43 & 1 & 0.5099 &  & Status:Bias:Group:Order & 21.44 & 10 & 0.0182 \\
Bias:Order & 10.49 & 5 & 0.0625 &  & Resume:Status:Bias:Group:Order & 9.91 & 10 & 0.4481 \\ \cline{6-9} 
Group:Order & 1.09 & 2 & 0.5812 &  &  &  &  &  \\ \cline{1-4}
\end{tabular}
\caption{Complete omnibus ANOVA results for RQ2 and RQ3.}
\end{table}

\begin{longtable}{rcccc}

\toprule

\multicolumn{5}{c}{\textbf{Model Fit}}  \\ 
\midrule
\multicolumn{5}{l}{\begin{tabular}[c]{@{}l@{}}Formula: Duration $\sim$Resume $\times$ Status $\times$ Bias $\times$ Group $\times$ Order +  Position + (1 | Participant) + (1 | Occupation)\end{tabular}} \\
\multicolumn{5}{l}{\begin{tabular}[c]{@{}l@{}}Standardized parameters were obtained by fitting the model on a standardized version of the dataset. \\ 95\% confidence intervals and p-values were computed using a Wald z-distribution approximation.\end{tabular}}  \\ 
\bottomrule

\multicolumn{1}{l}{} &  &  &  &   \\ 

\toprule
\multicolumn{5}{c}{\textbf{Random Effects}}\\
\multicolumn{1}{l}{} &  &  & Variance & SD \\ 
\midrule
\multicolumn{1}{l}{Participant} &  &  & 0.358 & 0.599  \\
\multicolumn{1}{l}{Occupation} &  &  & 0.002 & 0.050  \\
\bottomrule

\multicolumn{1}{l}{} &  &  &  &  \\ 

\toprule
\multicolumn{5}{c}{\textbf{Fixed Effects}} \\
 & Estimate & Std. Error & z value & Pr($>$\$|$z$|\$)  \\ 
\midrule
\endfirsthead

{{\tablename\ \thetable{} -- continued from previous page}} \\
\toprule
\multicolumn{5}{c}{\textbf{Fixed Effects}}\\
 & Estimate & Std. Error & z value & Pr($>$\$|$z$|\$) \\ 
\midrule
\endhead

\midrule
\multicolumn{3}{r}{Continued on next page} \\
\midrule
\caption{Regression fit information for RQ2 and RQ3.}
\endfoot

\bottomrule
\caption{Regression fit information for RQ2 and RQ3.}
\endlastfoot

(Intercept) & 3.22 & 0.03 & 98.71 & 0.00 \\ 
  Resume1 & -0.03 & 0.01 & -2.12 & 0.03 \\ 
  Status1 & 0.04 & 0.03 & 1.20 & 0.23 \\ 
  Bias1 & 0.34 & 0.02 & 16.66 & 0.00 \\ 
  Bias2 & -0.10 & 0.02 & -4.41 & 0.00 \\ 
  Bias3 & -0.06 & 0.04 & -1.60 & 0.11 \\ 
  Bias4 & -0.07 & 0.04 & -1.70 & 0.09 \\ 
  Bias5 & -0.09 & 0.03 & -3.00 & 0.00 \\ 
  Group1 & -0.03 & 0.04 & -0.81 & 0.42 \\ 
  Group2 & 0.04 & 0.04 & 0.88 & 0.38 \\ 
  Order1 & -0.01 & 0.03 & -0.28 & 0.78 \\ 
  Position1 & 0.29 & 0.02 & 14.06 & 0.00 \\ 
  Position2 & 0.09 & 0.02 & 4.52 & 0.00 \\ 
  Position3 & -0.03 & 0.02 & -1.57 & 0.12 \\ 
  Position4 & -0.15 & 0.02 & -7.24 & 0.00 \\ 
  Resume1:Status1 & -0.02 & 0.01 & -1.49 & 0.14 \\ 
  Resume1:Bias1 & 0.03 & 0.02 & 1.46 & 0.15 \\ 
  Resume1:Bias2 & -0.02 & 0.02 & -0.98 & 0.33 \\ 
  Resume1:Bias3 & 0.04 & 0.03 & 1.06 & 0.29 \\ 
  Resume1:Bias4 & -0.06 & 0.03 & -1.74 & 0.08 \\ 
  Resume1:Bias5 & 0.02 & 0.03 & 0.85 & 0.39 \\ 
  Status1:Bias1 & 0.00 & 0.02 & 0.02 & 0.99 \\ 
  Status1:Bias2 & -0.01 & 0.02 & -0.36 & 0.72 \\ 
  Status1:Bias3 & -0.03 & 0.04 & -0.76 & 0.44 \\ 
  Status1:Bias4 & 0.02 & 0.04 & 0.50 & 0.62 \\ 
  Status1:Bias5 & -0.01 & 0.03 & -0.31 & 0.76 \\ 
  Resume1:Group1 & 0.03 & 0.02 & 1.53 & 0.13 \\ 
  Resume1:Group2 & -0.03 & 0.02 & -1.70 & 0.09 \\ 
  Status1:Group1 & 0.00 & 0.04 & 0.05 & 0.96 \\ 
  Status1:Group2 & -0.02 & 0.04 & -0.57 & 0.57 \\ 
  Bias1:Group1 & -0.01 & 0.03 & -0.39 & 0.69 \\ 
  Bias2:Group1 & 0.01 & 0.03 & 0.21 & 0.83 \\ 
  Bias3:Group1 & 0.09 & 0.05 & 1.65 & 0.10 \\ 
  Bias4:Group1 & -0.00 & 0.05 & -0.00 & 1.00 \\ 
  Bias5:Group1 & -0.07 & 0.04 & -1.74 & 0.08 \\ 
  Bias1:Group2 & -0.03 & 0.03 & -0.98 & 0.33 \\ 
  Bias2:Group2 & -0.06 & 0.03 & -1.96 & 0.05 \\ 
  Bias3:Group2 & 0.05 & 0.05 & 1.00 & 0.32 \\ 
  Bias4:Group2 & -0.01 & 0.05 & -0.13 & 0.90 \\ 
  Bias5:Group2 & 0.06 & 0.04 & 1.50 & 0.13 \\ 
  Resume1:Order1 & -0.00 & 0.01 & -0.21 & 0.84 \\ 
  Status1:Order1 & 0.02 & 0.03 & 0.66 & 0.51 \\ 
  Bias1:Order1 & 0.05 & 0.02 & 2.32 & 0.02 \\ 
  Bias2:Order1 & -0.05 & 0.02 & -1.96 & 0.05 \\ 
  Bias3:Order1 & -0.02 & 0.04 & -0.60 & 0.55 \\ 
  Bias4:Order1 & -0.02 & 0.04 & -0.62 & 0.54 \\ 
  Bias5:Order1 & 0.02 & 0.03 & 0.63 & 0.53 \\ 
  Group1:Order1 & 0.03 & 0.04 & 0.60 & 0.55 \\ 
  Group2:Order1 & -0.04 & 0.04 & -1.03 & 0.30 \\ 
  Resume1:Status1:Bias1 & 0.01 & 0.02 & 0.47 & 0.64 \\ 
  Resume1:Status1:Bias2 & 0.03 & 0.02 & 1.21 & 0.23 \\ 
  Resume1:Status1:Bias3 & -0.03 & 0.03 & -0.85 & 0.39 \\ 
  Resume1:Status1:Bias4 & -0.01 & 0.03 & -0.37 & 0.71 \\ 
  Resume1:Status1:Bias5 & 0.07 & 0.03 & 2.77 & 0.01 \\ 
  Resume1:Status1:Group1 & -0.01 & 0.02 & -0.70 & 0.48 \\ 
  Resume1:Status1:Group2 & -0.02 & 0.02 & -1.23 & 0.22 \\ 
  Resume1:Bias1:Group1 & -0.02 & 0.03 & -0.82 & 0.41 \\ 
  Resume1:Bias2:Group1 & 0.02 & 0.03 & 0.50 & 0.62 \\ 
  Resume1:Bias3:Group1 & 0.08 & 0.05 & 1.68 & 0.09 \\ 
  Resume1:Bias4:Group1 & -0.02 & 0.05 & -0.46 & 0.65 \\ 
  Resume1:Bias5:Group1 & 0.00 & 0.04 & 0.03 & 0.97 \\ 
  Resume1:Bias1:Group2 & 0.01 & 0.03 & 0.38 & 0.70 \\ 
  Resume1:Bias2:Group2 & 0.04 & 0.03 & 1.30 & 0.19 \\ 
  Resume1:Bias3:Group2 & -0.08 & 0.05 & -1.70 & 0.09 \\ 
  Resume1:Bias4:Group2 & -0.02 & 0.05 & -0.40 & 0.69 \\ 
  Resume1:Bias5:Group2 & 0.03 & 0.03 & 0.78 & 0.43 \\ 
  Status1:Bias1:Group1 & -0.07 & 0.03 & -2.44 & 0.01 \\ 
  Status1:Bias2:Group1 & -0.02 & 0.03 & -0.58 & 0.56 \\ 
  Status1:Bias3:Group1 & 0.02 & 0.05 & 0.29 & 0.77 \\ 
  Status1:Bias4:Group1 & -0.07 & 0.05 & -1.34 & 0.18 \\ 
  Status1:Bias5:Group1 & 0.06 & 0.04 & 1.41 & 0.16 \\ 
  Status1:Bias1:Group2 & 0.07 & 0.03 & 2.57 & 0.01 \\ 
  Status1:Bias2:Group2 & -0.01 & 0.03 & -0.36 & 0.72 \\ 
  Status1:Bias3:Group2 & 0.06 & 0.05 & 1.07 & 0.29 \\ 
  Status1:Bias4:Group2 & -0.02 & 0.05 & -0.47 & 0.64 \\ 
  Status1:Bias5:Group2 & -0.07 & 0.04 & -1.89 & 0.06 \\ 
  Resume1:Status1:Order1 & -0.02 & 0.01 & -1.95 & 0.05 \\ 
  Resume1:Bias1:Order1 & 0.00 & 0.02 & 0.16 & 0.87 \\ 
  Resume1:Bias2:Order1 & -0.05 & 0.02 & -2.29 & 0.02 \\ 
  Resume1:Bias3:Order1 & -0.01 & 0.03 & -0.24 & 0.81 \\ 
  Resume1:Bias4:Order1 & -0.00 & 0.03 & -0.11 & 0.91 \\ 
  Resume1:Bias5:Order1 & 0.03 & 0.03 & 1.17 & 0.24 \\ 
  Status1:Bias1:Order1 & 0.02 & 0.02 & 0.94 & 0.34 \\ 
  Status1:Bias2:Order1 & -0.02 & 0.02 & -0.93 & 0.35 \\ 
  Status1:Bias3:Order1 & 0.08 & 0.04 & 2.19 & 0.03 \\ 
  Status1:Bias4:Order1 & 0.00 & 0.04 & 0.09 & 0.93 \\ 
  Status1:Bias5:Order1 & 0.02 & 0.03 & 0.58 & 0.56 \\ 
  Resume1:Group1:Order1 & -0.00 & 0.02 & -0.02 & 0.99 \\ 
  Resume1:Group2:Order1 & -0.01 & 0.02 & -0.76 & 0.45 \\ 
  Status1:Group1:Order1 & -0.00 & 0.04 & -0.01 & 0.99 \\ 
  Status1:Group2:Order1 & 0.08 & 0.04 & 1.80 & 0.07 \\ 
  Bias1:Group1:Order1 & -0.03 & 0.03 & -0.89 & 0.37 \\ 
  Bias2:Group1:Order1 & 0.08 & 0.03 & 2.34 & 0.02 \\ 
  Bias3:Group1:Order1 & 0.04 & 0.05 & 0.68 & 0.49 \\ 
  Bias4:Group1:Order1 & -0.01 & 0.05 & -0.17 & 0.86 \\ 
  Bias5:Group1:Order1 & 0.03 & 0.04 & 0.71 & 0.47 \\ 
  Bias1:Group2:Order1 & 0.01 & 0.03 & 0.32 & 0.75 \\ 
  Bias2:Group2:Order1 & -0.01 & 0.03 & -0.22 & 0.83 \\ 
  Bias3:Group2:Order1 & 0.04 & 0.05 & 0.83 & 0.40 \\ 
  Bias4:Group2:Order1 & -0.02 & 0.05 & -0.31 & 0.75 \\ 
  Bias5:Group2:Order1 & -0.02 & 0.04 & -0.40 & 0.69 \\ 
  Resume1:Status1:Bias1:Group1 & 0.03 & 0.03 & 1.03 & 0.30 \\ 
  Resume1:Status1:Bias2:Group1 & 0.02 & 0.03 & 0.49 & 0.62 \\ 
  Resume1:Status1:Bias3:Group1 & -0.10 & 0.05 & -2.25 & 0.02 \\ 
  Resume1:Status1:Bias4:Group1 & 0.03 & 0.05 & 0.55 & 0.58 \\ 
  Resume1:Status1:Bias5:Group1 & 0.08 & 0.04 & 2.25 & 0.02 \\ 
  Resume1:Status1:Bias1:Group2 & -0.02 & 0.03 & -0.88 & 0.38 \\ 
  Resume1:Status1:Bias2:Group2 & -0.01 & 0.03 & -0.37 & 0.71 \\ 
  Resume1:Status1:Bias3:Group2 & 0.02 & 0.05 & 0.37 & 0.71 \\ 
  Resume1:Status1:Bias4:Group2 & -0.01 & 0.05 & -0.28 & 0.78 \\ 
  Resume1:Status1:Bias5:Group2 & -0.02 & 0.03 & -0.51 & 0.61 \\ 
  Resume1:Status1:Bias1:Order1 & 0.01 & 0.02 & 0.48 & 0.63 \\ 
  Resume1:Status1:Bias2:Order1 & 0.00 & 0.02 & 0.03 & 0.98 \\ 
  Resume1:Status1:Bias3:Order1 & -0.07 & 0.03 & -2.10 & 0.04 \\ 
  Resume1:Status1:Bias4:Order1 & 0.05 & 0.03 & 1.37 & 0.17 \\ 
  Resume1:Status1:Bias5:Order1 & -0.02 & 0.03 & -0.85 & 0.40 \\ 
  Resume1:Status1:Group1:Order1 & 0.04 & 0.02 & 2.34 & 0.02 \\ 
  Resume1:Status1:Group2:Order1 & -0.04 & 0.02 & -2.15 & 0.03 \\ 
  Resume1:Bias1:Group1:Order1 & -0.01 & 0.03 & -0.57 & 0.57 \\ 
  Resume1:Bias2:Group1:Order1 & -0.00 & 0.03 & -0.15 & 0.88 \\ 
  Resume1:Bias3:Group1:Order1 & 0.05 & 0.05 & 1.01 & 0.31 \\ 
  Resume1:Bias4:Group1:Order1 & 0.03 & 0.05 & 0.58 & 0.56 \\ 
  Resume1:Bias5:Group1:Order1 & -0.01 & 0.04 & -0.35 & 0.72 \\ 
  Resume1:Bias1:Group2:Order1 & 0.03 & 0.03 & 1.16 & 0.24 \\ 
  Resume1:Bias2:Group2:Order1 & 0.03 & 0.03 & 1.13 & 0.26 \\ 
  Resume1:Bias3:Group2:Order1 & -0.01 & 0.05 & -0.14 & 0.89 \\ 
  Resume1:Bias4:Group2:Order1 & -0.06 & 0.05 & -1.23 & 0.22 \\ 
  Resume1:Bias5:Group2:Order1 & -0.06 & 0.03 & -1.83 & 0.07 \\ 
  Status1:Bias1:Group1:Order1 & 0.06 & 0.03 & 2.12 & 0.03 \\ 
  Status1:Bias2:Group1:Order1 & 0.06 & 0.03 & 1.81 & 0.07 \\ 
  Status1:Bias3:Group1:Order1 & -0.05 & 0.05 & -0.91 & 0.36 \\ 
  Status1:Bias4:Group1:Order1 & -0.10 & 0.05 & -1.88 & 0.06 \\ 
  Status1:Bias5:Group1:Order1 & 0.02 & 0.04 & 0.42 & 0.67 \\ 
  Status1:Bias1:Group2:Order1 & -0.11 & 0.03 & -3.73 & 0.00 \\ 
  Status1:Bias2:Group2:Order1 & -0.06 & 0.03 & -1.86 & 0.06 \\ 
  Status1:Bias3:Group2:Order1 & 0.03 & 0.05 & 0.57 & 0.57 \\ 
  Status1:Bias4:Group2:Order1 & 0.10 & 0.05 & 1.82 & 0.07 \\ 
  Status1:Bias5:Group2:Order1 & -0.00 & 0.04 & -0.06 & 0.95 \\ 
  Resume1:Status1:Bias1:Group1:Order1 & -0.03 & 0.03 & -1.23 & 0.22 \\ 
  Resume1:Status1:Bias2:Group1:Order1 & -0.06 & 0.03 & -1.88 & 0.06 \\ 
  Resume1:Status1:Bias3:Group1:Order1 & 0.07 & 0.05 & 1.54 & 0.12 \\ 
  Resume1:Status1:Bias4:Group1:Order1 & 0.04 & 0.05 & 0.91 & 0.36 \\ 
  Resume1:Status1:Bias5:Group1:Order1 & -0.00 & 0.04 & -0.01 & 0.99 \\ 
  Resume1:Status1:Bias1:Group2:Order1 & 0.01 & 0.03 & 0.50 & 0.61 \\ 
  Resume1:Status1:Bias2:Group2:Order1 & 0.01 & 0.03 & 0.46 & 0.64 \\ 
  Resume1:Status1:Bias3:Group2:Order1 & 0.02 & 0.05 & 0.44 & 0.66 \\ 
  Resume1:Status1:Bias4:Group2:Order1 & -0.07 & 0.05 & -1.50 & 0.13 \\ 
  Resume1:Status1:Bias5:Group2:Order1 & 0.03 & 0.03 & 0.91 & 0.36 \\ 

\end{longtable}

\section{RQ2 and RQ4 Omnibus ANOVA and Regression Table} \label{app:f}
\begin{table}[h!]
\begin{tabular}{lrrrllrrr}
\cline{1-4} \cline{6-9}
 & Chisq & Df & Pr($>$Chisq) &  &  & Chisq & Df & Pr($>$Chisq) \\ \cline{1-4} \cline{6-9} 
(Intercept) & 5574.20 & 1 & 0.0000 &  & Resume:Status:Bias & 4.83 & 5 & 0.4373 \\
Resume & 3.38 & 1 & 0.0660 &  & Resume:Status:Group & 3.39 & 2 & 0.1833 \\
Status & 1.78 & 1 & 0.1816 &  & Resume:Bias:Group & 16.53 & 10 & 0.0853 \\
Bias & 152.23 & 5 & 0.0000 &  & Status:Bias:Group & 31.23 & 10 & 0.0005 \\
Group & 3.01 & 2 & 0.2226 &  & Resume:Status:D & 0.04 & 1 & 0.8433 \\
D & 3.57 & 1 & 0.0587 &  & Resume:Bias:D & 1.90 & 5 & 0.8629 \\
Position & 294.05 & 4 & 0.0000 &  & Status:Bias:D & 2.88 & 5 & 0.7189 \\
Resume:Status & 0.87 & 1 & 0.3521 &  & Resume:Group:D & 6.09 & 2 & 0.0476 \\
Resume:Bias & 1.90 & 5 & 0.8627 &  & Status:Group:D & 3.30 & 2 & 0.1922 \\
Status:Bias & 4.59 & 5 & 0.4684 &  & Bias:Group:D & 20.85 & 10 & 0.0222 \\
Resume:Group & 8.10 & 2 & 0.0175 &  & Resume:Status:Bias:Group & 13.07 & 10 & 0.2196 \\
Status:Group & 2.62 & 2 & 0.2696 &  & Resume:Status:Bias:D & 2.36 & 5 & 0.7967 \\
Bias:Group & 16.11 & 10 & 0.0964 &  & Resume:Status:Group:D & 0.40 & 2 & 0.8202 \\
Resume:D & 0.01 & 1 & 0.9271 &  & Resume:Bias:Group:D & 19.27 & 10 & 0.0370 \\
Status:D & 0.35 & 1 & 0.5528 &  & Status:Bias:Group:D & 16.64 & 10 & 0.0827 \\
Bias:D & 7.45 & 5 & 0.1893 &  & Resume:Status:Bias:Group:D & 13.02 & 10 & 0.2227 \\ \cline{6-9} 
Group:D & 6.84 & 2 & 0.0328 &  &  &  &  &  \\ \cline{1-4}
\end{tabular}
\caption{Complete ombinus ANOVA results for RQ2 and RQ4.}   
\end{table}

\begin{longtable}{rcccc}

\toprule

\multicolumn{5}{c}{\textbf{Model Fit}}  \\ 
\midrule
\multicolumn{5}{l}{\begin{tabular}[c]{@{}l@{}}Formula: Duration $\sim$ Resume $\times$ Status $\times$ Bias $\times$ Group $\times$ D +  Position + (1 | Participant) + (1 | Occupation)\end{tabular}} \\
\multicolumn{5}{l}{\begin{tabular}[c]{@{}l@{}}Standardized parameters were obtained by fitting the model on a standardized version of the dataset. \\ 95\% confidence intervals and p-values were computed using a Wald z-distribution approximation.\end{tabular}}  \\ 
\bottomrule

\multicolumn{1}{l}{} &  &  &  &   \\ 

\toprule
\multicolumn{5}{c}{\textbf{Random Effects}}\\
\multicolumn{1}{l}{} &  &  & Variance & SD \\ 
\midrule
\multicolumn{1}{l}{Participant} &  &  & 0.350 & 0.591  \\
\multicolumn{1}{l}{Occupation} &  &  & 0.002 & 0.051  \\
\bottomrule

\multicolumn{1}{l}{} &  &  &  &  \\ 

\toprule
\multicolumn{5}{c}{\textbf{Fixed Effects}} \\
 & Estimate & Std. Error & z value & Pr($>$\$|$z$|\$)  \\ 
\midrule
\endfirsthead

{{\tablename\ \thetable{} -- continued from previous page}} \\
\toprule
\multicolumn{5}{c}{\textbf{Fixed Effects}}\\
 & Estimate & Std. Error & z value & Pr($>$\$|$z$|\$) \\ 
\midrule
\endhead

\midrule
\multicolumn{3}{r}{Continued on next page} \\
\midrule
\caption{Regression fit information for RQ2 and RQ4.}
\endfoot

\bottomrule
\caption{Regression fit information for RQ2 and RQ4.}
\endlastfoot

(Intercept) & 3.18 & 0.04 & 74.66 & 0.00 \\ 
  Resume1 & -0.03 & 0.02 & -1.84 & 0.07 \\ 
  Status1 & 0.06 & 0.04 & 1.34 & 0.18 \\ 
  Bias1 & 0.33 & 0.03 & 11.26 & 0.00 \\ 
  Bias2 & -0.12 & 0.04 & -3.47 & 0.00 \\ 
  Bias3 & 0.00 & 0.06 & 0.04 & 0.97 \\ 
  Bias4 & -0.04 & 0.06 & -0.71 & 0.48 \\ 
  Bias5 & -0.16 & 0.04 & -3.71 & 0.00 \\ 
  Group1 & -0.09 & 0.06 & -1.68 & 0.09 \\ 
  Group2 & 0.01 & 0.05 & 0.18 & 0.85 \\ 
  D & 0.13 & 0.07 & 1.89 & 0.06 \\ 
  Position1 & 0.29 & 0.02 & 14.03 & 0.00 \\ 
  Position2 & 0.10 & 0.02 & 4.61 & 0.00 \\ 
  Position3 & -0.02 & 0.02 & -1.18 & 0.24 \\ 
  Position4 & -0.15 & 0.02 & -7.30 & 0.00 \\ 
  Resume1:Status1 & -0.02 & 0.02 & -0.93 & 0.35 \\ 
  Resume1:Bias1 & 0.02 & 0.03 & 0.64 & 0.52 \\ 
  Resume1:Bias2 & -0.00 & 0.03 & -0.12 & 0.91 \\ 
  Resume1:Bias3 & 0.03 & 0.05 & 0.59 & 0.56 \\ 
  Resume1:Bias4 & -0.06 & 0.05 & -1.20 & 0.23 \\ 
  Resume1:Bias5 & 0.02 & 0.04 & 0.44 & 0.66 \\ 
  Status1:Bias1 & -0.01 & 0.03 & -0.45 & 0.65 \\ 
  Status1:Bias2 & -0.02 & 0.04 & -0.43 & 0.67 \\ 
  Status1:Bias3 & -0.08 & 0.06 & -1.50 & 0.13 \\ 
  Status1:Bias4 & 0.03 & 0.06 & 0.56 & 0.57 \\ 
  Status1:Bias5 & 0.01 & 0.04 & 0.28 & 0.78 \\ 
  Resume1:Group1 & 0.07 & 0.02 & 2.78 & 0.01 \\ 
  Resume1:Group2 & -0.03 & 0.02 & -1.54 & 0.12 \\ 
  Status1:Group1 & -0.06 & 0.06 & -1.15 & 0.25 \\ 
  Status1:Group2 & -0.04 & 0.05 & -0.66 & 0.51 \\ 
  Bias1:Group1 & -0.04 & 0.04 & -1.03 & 0.30 \\ 
  Bias2:Group1 & 0.00 & 0.05 & 0.08 & 0.93 \\ 
  Bias3:Group1 & -0.04 & 0.08 & -0.51 & 0.61 \\ 
  Bias4:Group1 & 0.06 & 0.08 & 0.82 & 0.41 \\ 
  Bias5:Group1 & -0.04 & 0.06 & -0.76 & 0.45 \\ 
  Bias1:Group2 & -0.03 & 0.04 & -0.77 & 0.44 \\ 
  Bias2:Group2 & -0.05 & 0.04 & -1.18 & 0.24 \\ 
  Bias3:Group2 & 0.07 & 0.07 & 1.02 & 0.31 \\ 
  Bias4:Group2 & -0.10 & 0.07 & -1.29 & 0.20 \\ 
  Bias5:Group2 & 0.18 & 0.05 & 3.38 & 0.00 \\ 
  Resume1:D & 0.00 & 0.03 & 0.09 & 0.93 \\ 
  Status1:D & -0.04 & 0.07 & -0.59 & 0.55 \\ 
  Bias1:D & 0.06 & 0.05 & 1.20 & 0.23 \\ 
  Bias2:D & 0.06 & 0.06 & 1.16 & 0.25 \\ 
  Bias3:D & -0.16 & 0.09 & -1.82 & 0.07 \\ 
  Bias4:D & -0.08 & 0.10 & -0.82 & 0.41 \\ 
  Bias5:D & 0.11 & 0.07 & 1.68 & 0.09 \\ 
  Group1:D & 0.11 & 0.09 & 1.25 & 0.21 \\ 
  Group2:D & 0.14 & 0.09 & 1.55 & 0.12 \\ 
  Resume1:Status1:Bias1 & 0.00 & 0.03 & 0.16 & 0.87 \\ 
  Resume1:Status1:Bias2 & 0.03 & 0.03 & 1.08 & 0.28 \\ 
  Resume1:Status1:Bias3 & -0.01 & 0.05 & -0.11 & 0.91 \\ 
  Resume1:Status1:Bias4 & 0.01 & 0.05 & 0.26 & 0.80 \\ 
  Resume1:Status1:Bias5 & 0.02 & 0.04 & 0.61 & 0.54 \\ 
  Resume1:Status1:Group1 & -0.01 & 0.02 & -0.42 & 0.68 \\ 
  Resume1:Status1:Group2 & -0.03 & 0.02 & -1.53 & 0.13 \\ 
  Resume1:Bias1:Group1 & -0.03 & 0.04 & -0.97 & 0.33 \\ 
  Resume1:Bias2:Group1 & -0.01 & 0.04 & -0.28 & 0.78 \\ 
  Resume1:Bias3:Group1 & 0.21 & 0.07 & 3.00 & 0.00 \\ 
  Resume1:Bias4:Group1 & 0.03 & 0.06 & 0.52 & 0.60 \\ 
  Resume1:Bias5:Group1 & -0.06 & 0.05 & -1.17 & 0.24 \\ 
  Resume1:Bias1:Group2 & 0.03 & 0.03 & 0.90 & 0.37 \\ 
  Resume1:Bias2:Group2 & -0.01 & 0.04 & -0.24 & 0.81 \\ 
  Resume1:Bias3:Group2 & -0.11 & 0.06 & -1.89 & 0.06 \\ 
  Resume1:Bias4:Group2 & -0.04 & 0.06 & -0.65 & 0.51 \\ 
  Resume1:Bias5:Group2 & 0.07 & 0.05 & 1.39 & 0.16 \\ 
  Status1:Bias1:Group1 & -0.05 & 0.04 & -1.22 & 0.22 \\ 
  Status1:Bias2:Group1 & -0.03 & 0.05 & -0.56 & 0.58 \\ 
  Status1:Bias3:Group1 & -0.06 & 0.08 & -0.73 & 0.47 \\ 
  Status1:Bias4:Group1 & -0.10 & 0.08 & -1.33 & 0.18 \\ 
  Status1:Bias5:Group1 & 0.04 & 0.06 & 0.69 & 0.49 \\ 
  Status1:Bias1:Group2 & 0.09 & 0.04 & 2.44 & 0.01 \\ 
  Status1:Bias2:Group2 & -0.01 & 0.04 & -0.27 & 0.78 \\ 
  Status1:Bias3:Group2 & 0.13 & 0.07 & 1.83 & 0.07 \\ 
  Status1:Bias4:Group2 & 0.02 & 0.07 & 0.31 & 0.76 \\ 
  Status1:Bias5:Group2 & -0.11 & 0.05 & -2.00 & 0.05 \\ 
  Resume1:Status1:D & -0.01 & 0.03 & -0.20 & 0.84 \\ 
  Resume1:Bias1:D & 0.03 & 0.04 & 0.82 & 0.41 \\ 
  Resume1:Bias2:D & -0.00 & 0.05 & -0.02 & 0.98 \\ 
  Resume1:Bias3:D & 0.06 & 0.08 & 0.77 & 0.44 \\ 
  Resume1:Bias4:D & -0.09 & 0.08 & -1.06 & 0.29 \\ 
  Resume1:Bias5:D & -0.00 & 0.06 & -0.03 & 0.97 \\ 
  Status1:Bias1:D & 0.00 & 0.05 & 0.06 & 0.95 \\ 
  Status1:Bias2:D & 0.03 & 0.06 & 0.49 & 0.63 \\ 
  Status1:Bias3:D & 0.11 & 0.09 & 1.20 & 0.23 \\ 
  Status1:Bias4:D & -0.03 & 0.10 & -0.29 & 0.77 \\ 
  Status1:Bias5:D & -0.02 & 0.07 & -0.30 & 0.76 \\ 
  Resume1:Group1:D & -0.10 & 0.04 & -2.38 & 0.02 \\ 
  Resume1:Group2:D & 0.02 & 0.04 & 0.44 & 0.66 \\ 
  Status1:Group1:D & 0.14 & 0.09 & 1.55 & 0.12 \\ 
  Status1:Group2:D & 0.02 & 0.09 & 0.20 & 0.84 \\ 
  Bias1:Group1:D & 0.06 & 0.07 & 0.86 & 0.39 \\ 
  Bias2:Group1:D & 0.02 & 0.08 & 0.26 & 0.80 \\ 
  Bias3:Group1:D & 0.29 & 0.13 & 2.19 & 0.03 \\ 
  Bias4:Group1:D & -0.20 & 0.14 & -1.46 & 0.14 \\ 
  Bias5:Group1:D & -0.00 & 0.09 & -0.02 & 0.98 \\ 
  Bias1:Group2:D & 0.01 & 0.06 & 0.10 & 0.92 \\ 
  Bias2:Group2:D & -0.05 & 0.07 & -0.61 & 0.54 \\ 
  Bias3:Group2:D & -0.08 & 0.12 & -0.68 & 0.50 \\ 
  Bias4:Group2:D & 0.25 & 0.13 & 1.91 & 0.06 \\ 
  Bias5:Group2:D & -0.28 & 0.09 & -3.11 & 0.00 \\ 
  Resume1:Status1:Bias1:Group1 & 0.03 & 0.04 & 0.80 & 0.43 \\ 
  Resume1:Status1:Bias2:Group1 & 0.04 & 0.04 & 0.83 & 0.41 \\ 
  Resume1:Status1:Bias3:Group1 & -0.09 & 0.07 & -1.22 & 0.22 \\ 
  Resume1:Status1:Bias4:Group1 & 0.04 & 0.06 & 0.58 & 0.56 \\ 
  Resume1:Status1:Bias5:Group1 & 0.04 & 0.05 & 0.80 & 0.42 \\ 
  Resume1:Status1:Bias1:Group2 & 0.01 & 0.03 & 0.26 & 0.79 \\ 
  Resume1:Status1:Bias2:Group2 & -0.00 & 0.04 & -0.09 & 0.92 \\ 
  Resume1:Status1:Bias3:Group2 & -0.02 & 0.06 & -0.31 & 0.76 \\ 
  Resume1:Status1:Bias4:Group2 & -0.13 & 0.06 & -2.12 & 0.03 \\ 
  Resume1:Status1:Bias5:Group2 & 0.04 & 0.05 & 0.94 & 0.35 \\ 
  Resume1:Status1:Bias1:D & 0.00 & 0.04 & 0.01 & 1.00 \\ 
  Resume1:Status1:Bias2:D & -0.02 & 0.05 & -0.45 & 0.65 \\ 
  Resume1:Status1:Bias3:D & -0.03 & 0.08 & -0.45 & 0.66 \\ 
  Resume1:Status1:Bias4:D & -0.03 & 0.08 & -0.38 & 0.71 \\ 
  Resume1:Status1:Bias5:D & 0.09 & 0.06 & 1.49 & 0.14 \\ 
  Resume1:Status1:Group1:D & 0.00 & 0.04 & 0.09 & 0.93 \\ 
  Resume1:Status1:Group2:D & 0.02 & 0.04 & 0.52 & 0.60 \\ 
  Resume1:Bias1:Group1:D & 0.02 & 0.06 & 0.41 & 0.69 \\ 
  Resume1:Bias2:Group1:D & 0.04 & 0.07 & 0.57 & 0.57 \\ 
  Resume1:Bias3:Group1:D & -0.28 & 0.12 & -2.39 & 0.02 \\ 
  Resume1:Bias4:Group1:D & -0.16 & 0.12 & -1.32 & 0.19 \\ 
  Resume1:Bias5:Group1:D & 0.16 & 0.08 & 1.84 & 0.07 \\ 
  Resume1:Bias1:Group2:D & -0.03 & 0.06 & -0.58 & 0.56 \\ 
  Resume1:Bias2:Group2:D & 0.12 & 0.07 & 1.66 & 0.10 \\ 
  Resume1:Bias3:Group2:D & 0.07 & 0.10 & 0.63 & 0.53 \\ 
  Resume1:Bias4:Group2:D & 0.10 & 0.11 & 0.90 & 0.37 \\ 
  Resume1:Bias5:Group2:D & -0.11 & 0.08 & -1.45 & 0.15 \\ 
  Status1:Bias1:Group1:D & -0.04 & 0.07 & -0.57 & 0.57 \\ 
  Status1:Bias2:Group1:D & 0.02 & 0.08 & 0.22 & 0.83 \\ 
  Status1:Bias3:Group1:D & 0.13 & 0.13 & 1.01 & 0.31 \\ 
  Status1:Bias4:Group1:D & 0.14 & 0.14 & 1.00 & 0.32 \\ 
  Status1:Bias5:Group1:D & 0.01 & 0.09 & 0.13 & 0.90 \\ 
  Status1:Bias1:Group2:D & -0.06 & 0.06 & -0.91 & 0.36 \\ 
  Status1:Bias2:Group2:D & -0.02 & 0.07 & -0.21 & 0.84 \\ 
  Status1:Bias3:Group2:D & -0.14 & 0.12 & -1.19 & 0.24 \\ 
  Status1:Bias4:Group2:D & -0.12 & 0.13 & -0.95 & 0.34 \\ 
  Status1:Bias5:Group2:D & 0.07 & 0.09 & 0.82 & 0.41 \\ 
  Resume1:Status1:Bias1:Group1:D & -0.01 & 0.06 & -0.11 & 0.91 \\ 
  Resume1:Status1:Bias2:Group1:D & -0.05 & 0.07 & -0.71 & 0.48 \\ 
  Resume1:Status1:Bias3:Group1:D & -0.01 & 0.12 & -0.12 & 0.90 \\ 
  Resume1:Status1:Bias4:Group1:D & -0.02 & 0.12 & -0.19 & 0.85 \\ 
  Resume1:Status1:Bias5:Group1:D & 0.09 & 0.08 & 1.01 & 0.31 \\ 
  Resume1:Status1:Bias1:Group2:D & -0.08 & 0.06 & -1.37 & 0.17 \\ 
  Resume1:Status1:Bias2:Group2:D & -0.02 & 0.07 & -0.27 & 0.79 \\ 
  Resume1:Status1:Bias3:Group2:D & 0.07 & 0.10 & 0.71 & 0.48 \\ 
  Resume1:Status1:Bias4:Group2:D & 0.29 & 0.11 & 2.62 & 0.01 \\ 
  Resume1:Status1:Bias5:Group2:D & -0.14 & 0.08 & -1.71 & 0.09 \\

\end{longtable}

\section{D and Demographic Trait Associations} \label{app:g}

We examined the associations between D and participants' self-reported age, gender, race, and state of residence. The correlation between D and age was low ($r=.049$). The other variables were non-numeric, so we conducted a Type III ANOVA to identify significant associations. There was no significant association between D and race ($F(14, 513) = 1.64$, $p = 0.065$) or state of residence ($F(45, 482) = 0.93$, $p = 0.610$). There was a significant but small association between D and gender ($F(4, 523) = 2.85$, $p = 0.023$). Post-hoc pairwise comparisons showed that significant differences in D scores existed between non-binary participants and men ($t(523)=-2.89$, $p=.004$), women ($t(523)=3.01$, $p=.003$), or those who selected more than one option ($t(523)=2.64$, $p=.008$). Because the sample size of non-binary participants in our data was small and we had no a priori hypotheses about the effects of gender in this study, we leave it to future research to determine how this may impact AI-HITL decision-making tasks.

\section{Scenario Ordering} \label{app:h}

\begin{figure}[h!]
    \centering
    \includesvg[width=0.6\textwidth]{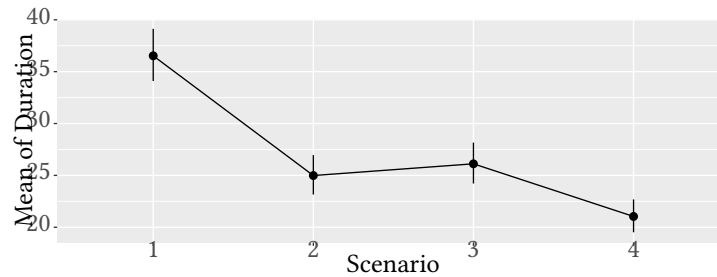}
    \caption{Estimated duration spent viewing resumes by participants' self-reported recruiting experience.}
    \label{fig:p_scenario}
\end{figure}

It has been widely observed that the time taken to complete cognitive tasks decreases according to a power law as the number of task repetitions increase, which is often known as a practice effect \citep{anderson2013cognitive}. This effect can be seen for this study in Figure \ref{fig:p_scenario}, where the pattern is broadly that participants spend more time viewing resumes in the scenario completed first (which has no AI recommendations) than in any of the scenarios completed later (which have AI recommendations). This introduces a confound where the decrease in time spent on scenarios with AI recommendations could be attributed to either practice effects (H1), the presence or absence of AI recommendations (H2), or a combination of the two (H3).

To examine these alternative hypotheses, we fit four zero-inflated general linear mixed models (Gamma family with a log link) (estimated using REML and nlminb optimizer) to predict Duration with variables representing scenario ordering that differentiate practice effects from effects due to our experimental manipulations. Each model also had random effects for Participant and Occupation. The first model is an agnostic baseline where a factor with four levels represents the positions scenarios occurred at within the experiment. The second model corresponds to H1, where the logarithm of the numeric positions of scenarios in the experiment is used to predict viewing duration. The third model recodes scenario positions into a new factor with two levels which represent H2: No AI (the first scenario) and AI (the second, third, and final scenarios). The final model corresponding to H3 includes the fixed effects from both the second and third models to assess whether a combination of the two explanations is superior to either in isolation.

We assessed the fit of each of these models using their Akaike information criteria (AIC) \citep{akaike1974new} and Bayesian information criterion (BIC) \citep{schwarz1978estimating}, which enable comparison of models that do not have hierarchical relationships \citep{vrieze2012model}. AIC favors models which best predict the data, and BIC favors model which predict the data with as few parameters as possible \citep{kuha2004aic}. The AIC and BIC results for each model are shown in Table \ref{tab:aic}, and for each metric lower values correspond to better models. For both AIC and BIC, the baseline model performs the best, followed by the hybrid model H3 for AIC and the logarithmic model H1 for BIC.

\begin{table}[]
\centering
\begin{tabular}{@{}lcc@{}}
\toprule
\multicolumn{1}{c}{\textbf{Model}} & \textbf{AIC} & \textbf{BIC} \\ \midrule
Baseline & 50956.8 & 51010.9 \\
H1 & 50973.8 & 51014.4 \\
H2 & 50984.6 & 51025.2 \\
H3 & 50970.2 & 51017.6 \\ \bottomrule
\end{tabular}
\caption{Information loss across models with different encoding schemes for scenario order to investigate the presence of practice effects.}
\label{tab:aic}
\end{table}

These results suggest that both whether a scenario has AI recommendations or not and its relative positioning in the experiment contain relevant information about the duration spent reading resumes in the hiring task ($AIC(H3) < AIC(H1) < AIC(H2)$). However, these two factors do not have all relevant information for predicting duration ($AIC(Baseline) < AIC(H3)$ and $BIC(Baseline) < BIC(H3)$). Furthermore, when greater model complexity is more heavily penalized, the logarithmic model performs better than the hybrid model ($BIC(H1) < BIC(H3) < BIC(H2)$). This follow-up analysis suggests that practice effects do play some role in explaining resume viewing duration; future experiments can test the extend of this more explicitly by conducting experiments in which the presentation order of blocks with and without AI recommendations are counterbalanced.

\section{Recruiter Experience} \label{app:i}

\begin{figure}[h!]
    \centering
    \includesvg[width=0.6\textwidth]{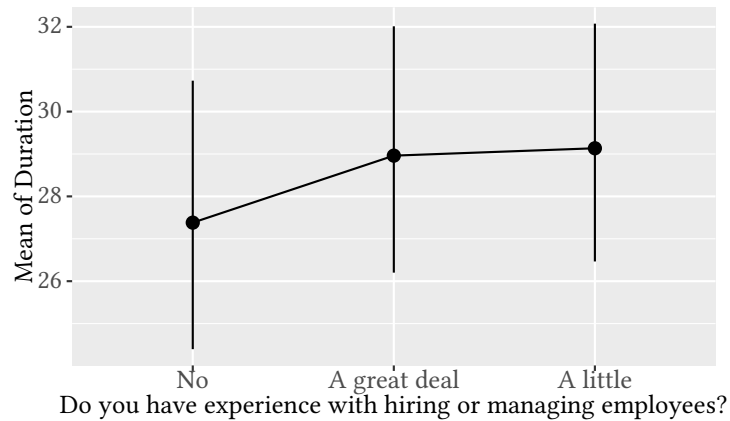}
    \caption{Estimated duration spent viewing resumes by participants' self-reported recruiting experience.}
    \label{fig:p_experience}
\end{figure}

As reported in \citet{wilson2025no}, the majority of participants reported having a little (24.8\%) or a great deal (38.4\%) of experience hiring and managing employees. Slightly more than one-third (36.7\%) of participants reporting having no hiring experience. To examine whether hiring experience (Experience) was associated with the amount of time participants spent viewing resumes, we fit another zero-inflated general linear mixed model (Gamma family with a log link) (estimated using REML and nlminb optimizer) to predict Duration with fixed effects for Experience and random effects for Participant and Occupation and tested omnibus significance with an ANOVA. The model's total explanatory power is substantial (conditional R2 = 0.43) and the part related to the fixed effects alone (marginal R2) is approximately 0.0008. The factor Experience is not significantly associated with Duration ($\chi^2(2) = 0.825$, $p = 0.662$), as shown in Figure \ref{fig:p_experience}.

\end{document}